\documentclass[onecolumn,pra, showpacs, superscriptaddress]{revtex4}
\usepackage{amsmath,amssymb,mathrsfs}

\usepackage{hyperref}
\usepackage{paralist}
\usepackage{pdfpages}
\usepackage{lscape} %
 \usepackage{graphicx}

\def\be{\begin{equation}}
\def\ee{\end{equation}}
\def\bea{\begin{eqnarray}}
\def\eea{\end{eqnarray}}
\def\nn{\nonumber}

\newcommand{\beq}{\begin{equation}}
\newcommand{\eeq}{\end{equation}}

\newcommand{\beqa}{\begin{eqnarray}}
\newcommand{\eeqa}{\end{eqnarray}}

\newcommand{\Beqa}{\begin{eqnarray*}}
\newcommand{\Eeqa}{\end{eqnarray*}}

\begin{document}

\title{ Momentum distribution  and contacts of   one-dimensional spinless Fermi gases
with an attractive p-wave interaction}

\author{Xiangguo Yin}
\email[e-mail:]{yinxiangguo@sxu.edu.cn}           
\affiliation{Institute of Theoretical Physics, Shanxi University, Taiyuan 030006, China}
\affiliation{Department of Physics and Center of Theoretical and Computational Physics, The University of Hong Kong, Hong Kong, China}    

\author{Xi-Wen Guan}
\address{State Key Laboratory of Magnetic Resonance and Atomic and Molecular Physics, Wuhan Institute of Physics and Mathematics, Chinese Academy of Sciences, Wuhan 430071, People's Republic of China} 
\address{Department of Theoretical Physics, Research School of Physics and Engineering, Australian National University, Canberra ACT 0200, Australia}       

\author{Yunbo Zhang}       
\affiliation{Institute of Theoretical Physics, Shanxi University, Taiyuan 030006, China}

\author{Haibin Su}
\affiliation{Department of Chemistry, The Hong Kong University of Science and Technology, Hong Kong, China}

\author{Shizhong Zhang}         
\affiliation{Department of Physics and Center of Theoretical and Computational Physics, The University of Hong Kong, Hong Kong, China}

\begin{abstract}
We present a rigorous  study of momentum distribution and  $p$-wave contacts of one dimensional (1D) spinless Fermi gases with an attractive  p-wave interaction. Using the Bethe wave function, we analytically calculate  the large-momentum tail of  momentum distribution of the model. We show that the  leading ($\sim 1/p^{2}$) and sub-leading terms ($\sim 1/p^{4}$) of the large-momentum tail are determined by two contacts $C_2$ and $C_4$, which we show, by explicit calculation, are related to the short-distance behaviour of the two-body correlation function and its derivatives. We show as one increases the 1D scattering length, the contact $C_2$ increases monotonically from zero while $C_4$ exhibits a peak for finite scattering length. In addition, we obtain analytic expressions for $p$-wave contacts at finite temperature from the thermodynamic Bethe ansatz equations in both weakly and strongly attractive regimes. 
\end{abstract}

\pacs{03.75.Ss, 03.75.Hh, 02.30.IK, 05.70.Ce}

\maketitle

\section{Introduction}

In the past few decades, experimental advances have made it possible to engineer with high controllability one-dimensional systems of ultracold atoms. Furthermore, the interactions between atoms can be tuned by a variety of experimental techniques, thus offering a promising opportunity to realize one-dimensional (1D) models of interacting spins, bosons and fermions.  It is well-known that ultra-cold atomic gases in 1D display a rich paradigm of few-body to many-body physics~\cite{CIR_Olshanii1998,CIR_Olshanii2003,exp_TG_Paredes2004, exp_TG_Kinoshita2004}. In contrast to the study of quantum many-body systems in higher dimensions,  many 1D systems can be treated in exact manners, such as Bethe ansatz approach~\cite{RevModPhys.83.1405,GBL:2013}, Bose-Fermi mapping~\cite{BF_mapping_Girardeau1960}, and quantum field theory~\cite{Giamarchi:2004}.

Exactly solvable models provide important benchmark understanding of quantum many-body phenomena, ranging from  quantum correlations to  quantum criticality and quantum liquids~\cite{Korepin,Takahashi-b,1D-Hubbard,Sutherland-book}. In this regard, the prototypical exactly solvable model of the Lieb-Liniger Bose gas provides  a deep understanding of quantum statistics, thermodynamics and quantum critical phenomena~\cite{BA_Lieb_Liniger1963, BA_Yang1967}, see a review \cite{Jiang:2015}. The Bethe ansatz solution of this  model is not only widely used to perform analytical calculations of important physical quantities  which shed light on universal behaviour of many-body systems, but also  presents a test ground to explore  equilibrium and nonequilibrium physics in experiment, for example,  Tonks-Girardeau gases \cite{exp_TG_Paredes2004, exp_TG_Kinoshita2004}, super Tonks-Girardeau gases \cite{exp_sTG_Haller2009}, quantum liquids \cite{Yang:2017}, thermalization of 1D ensemble of cold atoms \cite{Kinoshita2006, Hofferberth2007} etc. 

By using optical lattices, low dimensional quantum gases with rich internal degrees of freedom can be  realized, where the interaction range and scattering length can be tuned. Such advances in experiment stimulate wide theoretical interests in 1D quantum spinor gases with high spin symmetries and high partial waves interactions.  In strong coupling regime, these systems display quite different spin charge separations depending quantum statistics of constituent particles~\cite{GBT:2007,effective_spin_chain_Deuretzbacher2014, effective_spin_chain_Murmann2015, effective_spin_chain_Levinsen2015, effective_spin_chain_Puhan2015, effective_spin_chain_Chenshu2016, effective_spin_chain_Liuyanxia2017, effective_spin_chain_Panlei2017}.

A universal theme in the study of dilute atomic gas is the correspondence between two-body and many-body correlations at short distance~\cite{Tan2008_1, Tan2008_2, Tan2008_3, Zhang2009, Werner2009,contact_Braaten2009, exp_contact_Jin2010}. This correspondence manifests itself in the  relation between various physical quantities for which the so-called contact plays a central role. These relations include adiabatic sweep relation, the large-moment tail of momentum distribution,
the derivative of the free energy with respect to a scattering length, Virial theorem, pressure relation, and so on. A nice feature of these relations is that they apply to both bosons and fermions in any dimensions, irrespective of the states the system is in. 

Very recently, universal
relations in  contact are found in  systems of ultracold atoms with  a $p$-wave interaction~\cite{Contact_p_wave_Yu2015, contact_p_wave_Ueda2015, contact_p_wave_cui2016, contact_p_wave_cui2016_2, contact_p_wave_Peng2016, contact_p_wave_Zhou2016, exp_contact_p_wave_Luciuk2016}. For spinless Fermi gas, there is no $s$-wave interaction and $p$-wave interaction dominates. The p-wave interaction parameters, including the scattering volume
$v$ and the effective range $R$, can be controllable by a magnetic field or a confinement-induced-resonance. At low energies, the phase shift $\delta_{p}\left(k\right)$ of $p$-wave interaction can be expressed as 
\begin{equation}
k^{3}\cot\delta_{p}\left(k\right)=-\frac{1}{v}-\frac{k^{2}}{R}+\mathcal{O}\left(k^{4}\right)
\end{equation}
where $k$ is the relative momentum of colliding atoms. It is important to note that inclusion of effective range $R$ in the parametrisation of the low-energy scattering spoils the Bose-Fermi mapping with a spinless Bose gas.


Such a complexity of interactions impose a major theoretical challenge in studying quantum correlations at a man-body level. Therefore the study of the $p$-wave contacts of  interacting fermions through exactly solvable models is high desirable. In this paper,  we develop the Bethe ansatz wave function  method  to calculate  the large-momentum distribution of 1D spinless Fermi gas with an attractive p-wave interaction, and show  that the  leading and sub-leading terms ( $\sim 1/p^{2}$ and  $\sim 1/p^{4}$ ) of the  large-momentum tail   respectively give rise to  two contacts $C_2$ and $C_4$ which are determined by the two-body correlation function and its derivative. We obtain exact the large-momentum tail of momentum distribution in both the weak and strong interaction limits. Universal behaviour of the $p$-wave contact is discussed. 

The paper is organised as follows. In section II we discuss  the interaction
boundary condition and the Bethe ansatz  solution of of the 1D spinless
Fermi gases with a p-wave interaction. In Section III, we present  our  main
results of the p-wave contacts, the large-momentum asymptotic of
the momentum distribution. We find that the coefficients of the universal
leading $1/p^{2}$ and sub-leading $1/p^{4}$ tails are related to
the two-body correlation function and its derivative, respectively. 
In Section IV we calculate the two body correlation function and contacts
in weak-coupling regime, and test the relationship between correlation
function and energy derivatives. 
In addition we show the consistence between
the statistical approach and the thermodynamical Bethe-Ansatz method.
Section V discusses  the contacts in strong attractive regime. 
We conclude in Section VI.

\section{Model}
\label{sec2}

We consider a 1D system composed of spinless Fermi gas with a $p$-wave
interaction, which implies the following interaction
boundary condition for wavefunction $\psi\left(z_{1},\cdots,z_{N}\right)$ \cite{interaction_p_wave_Pricoupenko2008,BA_p_wave_Imambekov2010}
\begin{equation}
\lim_{z=z_{j}-z_{i}\rightarrow0^{+}}\left(\frac{1}{a_{p}^{\mathrm{1D}}}+\partial_{z}-R^{\mathrm{1D}}\partial_{z}^{2}\right)\psi\left(z_{1},\cdots,z_{N}\right)=0,\label{eq:interaction-boundary-condition}
\end{equation}
where $z_{1},\cdots,z_{N}$ are coordinates of $N$ fermions. 
Under a strong two-dimensional harmonic
confinement, only the lowest transverse
mode is occupied and the 1D scattering length
$a_{p}^{\mathrm{1D}}$ and 1D effective range $R^{\mathrm{1D}}$ are
related to the 3D scattering volume $v$ and 3D effective range $R$  through
\begin{equation}
a_{p}^{\mathrm{1D}}=3a_{\bot}\left[\frac{a_{\bot}^{3}}{v}-3\sqrt{2}\zeta\left(-1/2\right)\right]^{-1},\:R^{\mathrm{1D}}=\frac{a_{\bot}^{2}}{3R},\label{eq:a-1D}
\end{equation}
respectively. Here  $3\sqrt{2}\zeta\left(-1/2\right)\approx-0.88$. The above relations in (\ref{eq:a-1D}) requires that the momenta of scattering fermions to satisfy $ka_{\bot}\ll1$.  As usual, $a_{\bot}=\sqrt{\hbar/m\omega_{\bot}}$ is the transverse oscillator length, $m$ is  the atomic mass and $\omega_{\bot}$ is the trapping frequency. 

By using the  above boundary condition (\ref{eq:interaction-boundary-condition}) and the asymptotic Bethe ansatz,  the eigenvalues and eigenfunctions of the  uniform system have been exactly obtained~\cite{BA_p_wave_Hao2007, BA_p_wave_Imambekov2010}. The wave function in the domain $\left\{ 0\leq z_{1}\leq\cdots\leq z_{N}\leq L\right\}$ is given in terms of  the  superpositions  of $N!$ plane waves
\begin{equation}
\psi\left(z_{1},\cdots,z_{N}\right)=\sum_{P}a\left(P\right)\exp\left(\mathrm{i}\sum_{l=1}^{N}\lambda_{P_{l}}z_{l}\right),\label{eq:wave-function-ansatz}
\end{equation}
where  $\lambda_{1},\cdots,\lambda_{N}$ are quasi-momenta.
In the above equation,  the summation account for  all permutations $P$s of the $N$ numbers ${1},\, 2, \, \cdots,\,{N}$,
and $a\left(P\right)$ stands for the  coefficients depending on the
quasi-momenta,  $p$-wave interaction parameters, i.e. the scattering length  $a_{p}^{\mathrm{1D}}$
and the effective range  $R^{\mathrm{1D}}$. The wave function in other domains can be obtained  by the antisymmetry of exchanging any two atoms. 
%
%
The energy of system is given by  $E=\hbar^{2}/\left(2m\right)\sum_{i=1}^{N}\lambda_{i}^{2}$.
In the following calculation, we take the units $\hbar=2m=1$. 
Using the  interaction boundary
condition (\ref{eq:interaction-boundary-condition}) and the  wave function  with periodic boundary conditions (\ref{eq:wave-function-ansatz}), we can obtain 
the amplitudes 
\begin{equation}
a\left(P\right)=\left(-1\right)^{P}\prod_{i<j=1}^{N}\left(S\left(\lambda_{P_{i}}-\lambda_{P_{j}}\right)\right)^{1/2}, \label{eq:AP}
\end{equation}
where $S(x)=\frac{\xi_{p}x^{2}-1/{\left|l_{p}\right|}+\textrm{i}x}{\xi_{p}x^{2}-1/{\left|l_{p}\right|}-\textrm{i}x}$, and we denote the scattering length $l_{p}=a_{p}^{\mathrm{1D}}/2$ and the effective range $\xi_{p}=R^{\mathrm{1D}}/2$. Note that $\left(-1\right)^{P}=+1$ $\left(-1\right)$ for an even (odd) permutation. It follows that the BA equations read \cite{BA_p_wave_Imambekov2010}
\begin{equation}
\exp\left(\textrm{i}\lambda_{i}L\right)=\prod_{j=1}^{N}S\left(\lambda_{i}-\lambda_{j}\right),\:i=1,2,\cdots,N\label{eq:BA-equations}
\end{equation}
that  determine the value of quasi-momenta $\lambda_{1},\cdots,\lambda_{N}$. The BA equations  provide exact ground state and excitations of the 1D spinless fermions with a p-wave interactions. For $l_{p}=0$, the BA equations (\ref{eq:BA-equations}) naturally reduce to the quasimomenta of free fermions, for which the  wave function $\psi\left(z_{1},\cdots,z_{N}\right)$
is given by Slater determinant.
Whereas at p-wave resonance, i.e. $l_{p}=\infty$, the BA equations  (\ref{eq:BA-equations}) reduce to that of the Lieb-Liniger  Bose gas \cite{BA_Lieb_Liniger1963} with the coupling constant $c_B=1/\xi_p$. At the $p$-wave resonance, a large value of $\xi_p$ drives the $p$-wave spinless Fermi gas into the regime of the weakly interacting bosons.
This gives a very interesting physical regime and likely to be reachable in experiment. 
In contrust, for $\xi_p=0$, the BA equations (\ref{eq:BA-equations}) reduce to the ones for the 1D Lieb-Liniger Bose gas with the interacting strength $c=1/|l_p|$, reserving the Bose-Fermi mapping. 
The  instability of this model was  discussed in~\cite{Pan:2018}.
In this paper, we only focus  the case of  $a_{p}^{\mathrm{1D}}<0$ and $R^{\mathrm{1D}}>0$,
for which the solution of quasi-momenta are real. 
The general solutions to the BA equations  (\ref{eq:BA-equations}) are much more complicated. 
As usual,  by taking  logarithm of both sides of the equations  (\ref{eq:BA-equations}), the roots are determined by 
\begin{equation}
 \lambda_{i}L=2\pi J_{i}-\sum_{j=1}^{N}\theta\left(\lambda_{i}-\lambda_{j}\right), \label{eq:BA-equations2}
 \end{equation}
where  the phase shift  is given by $\theta\left(\lambda\right)=2\arg\left(\textrm{i}\lambda-\xi_{p}\lambda^{2}+1/\left|l_{p}\right|\right)$. 
In the above equations, the quantum numbers  $J_{i}$ take an integer (half an integer) for odd (even) particle number $N$.
The  thermodynamics of this model was presented in~\cite{Chen:2016}.

\section{III. Momentum Distribution}

Our object is the asymptotic behavior of the momentum distribution, which determines the $p$-wave contacts. The dominant contribution in the large-momentum tail of the momentum distribution involves the singular behaviour of the wave function in the vicinity of the interaction
point, i.e. $z_{ij}=z_{i}-z_{j} \rightarrow 0$. In order to evaluate it, here we generalize  the method for calculating the large-momentum distribution~\cite{Olshanii2003}  and the method for  calculating the multipartcle local correlation functions~\cite{correlation_Nandani2016}.
To this end, three major steps are needed: (a) Taylor series
expansion of the wave function in the vicinity of  the interaction point; 
 (b) the asymptotics of Fourier integral of the wave faction;
 and (c) two-body correlation functions.
 We will discuss the major  step   correlation function in the next Section.

{\bf (a) Taylor series expansion of the wave function.}
Without losing generality, we consider the interaction point of the first
and the $i$-th particles.
Taking into account of the antisymmetry of the wave function,  it can be  expanded in terms of   $z_{i1}\equiv z_1-z_j$
\begin{align}
\psi\left(z_{1},\cdots,z_{N}\right)=&\psi_{0}\left(Z_{i}\right)\textrm{sgn}\left(z_{i1}\right)+\psi_{1}\left(Z_{i}\right)z_{i1}+ \nn\\
&\psi_{2}\left(Z_{i}\right)\textrm{sgn}\left(z_{i1}\right)z_{i1}^{2}+\mathcal{O}\left(z_{i1}^{3}\right)\label{eq:wave-function-expansion}
\end{align}
with 
\begin{align}
\psi_{0}\left(Z_{i}\right) &=\lim_{\varepsilon\to 0^+}\left.\psi\left(z_{1},\cdots,z_{N}\right)\right|_{z_{1}=z_{i}-\varepsilon},\label{eq:phi0}\\
\psi_{1}\left(Z_{i}\right) &=\frac{1}{2}\lim_{\varepsilon\to 0^+}\left.\left[\left(\partial_{z_{i}}-\partial_{z_{1}}\right)\psi\left(z_{1},\cdots,z_{N}\right)\right]\right|_{z_{1}=z_{i}-\varepsilon},\label{eq:phi2}\\
\psi_{2}\left(Z_{i}\right)&=\frac{1}{8}\lim_{\varepsilon\to 0^+}\left.\left[\left(\partial_{z_{i}}-\partial_{z_{1}}\right)^{2}\psi\left(z_{1},\cdots,z_{N}\right)\right]\right|_{z_{1}=z_{i}-\varepsilon}.\label{eq:phi3}
\end{align}
Here $Z_{i1}=\left(z_{i}+z_{1}\right)/2$ and $z_{i1}=z_{i}-z_{1}$
are the center-of-mass and relative coordinates of the $(1i)$ pair
of particles, respectively.
We will denote by $Z_{i}=\left\{ Z_{i1},z_{2},\cdots,z_{i-1},z_{i+1},\cdots,z_{N}\right\}$ the center-of-mass coordinate of the first and $i$-th particles and the coordinates of all the rest of the particles. In the above equations the sign function is defined as $\textrm{sgn}\left(z\right)=-1$ for
$z<0$; $\textrm{sgn}\left(z\right)=0$, for $z=0$; and $\textrm{sgn}\left(z\right)=+1$
for $z>0$.
 Due to Pauli exclusion principle, the wave function is
zero at the interaction point, and it is not continuous owing to the $p$-wave
interaction for the 1D spinless Fermi gas with an attractive p-wave interaction. 
Here we calculate the  functions contributions from the terms involving the functions $\psi_{0}\left(Z_{i}\right)$,
$\psi_{1}\left(Z_{i}\right)$, $\psi_{2}\left(Z_{i}\right)$ at
position $z_{i}=z_{1}+\varepsilon$. 

{\bf (b)  The asymptotics of Fourier integral of the wave function.}  In general,  for the periodic functions $\mathrm{sgn}\left(z_{0}-z\right)F\left(z\right)$,
$\left(z_{0}-z\right)F\left(z\right)$, and $\mathrm{sgn}\left(z_{0}-z\right)\left(z_{0}-z\right)^{2}F\left(z\right)$
which are  defined on the interval $\left[0,L\right]$, where $F\left(z\right)$
is a regular function, we can directly calculate their Fourier transforms  through integration by parts.  Up
to the order of  $1/p^{3}$, we obtain  asymptotics  of the Fourier transforms of these 
functions 
\begin{align}
&\int_{0}^{L}dze^{-\mathrm{i}pz}\mathrm{sgn}\left(z_{0}-z\right)F\left(z\right)\nn\\
=&\left[\frac{2\mathrm{i}}{p}F\left(z\right)+\frac{2}{p^{2}}\partial_{z}F\left(z\right)-\frac{2\textrm{i}}{p^{3}}\partial_{z}^{2}F\left(z\right)\right]_{z=z_{0}}e^{-\mathrm{i}pz_{0}}+\mathcal{O}\left(\frac{1}{p^{4}}\right),\label{eq:asymptotics-0}
\end{align}
\begin{equation}
\int_{0}^{L}dze^{-\mathrm{i}pz}\left(z_{0}-z\right)F\left(z\right)=0,\label{eq:asymptotics-1}
\end{equation}
\begin{align}
&\int_{0}^{L}dze^{-\mathrm{i}pz}\mathrm{sgn}\left(z_{0}-z\right)\left(z_{0}-z\right)^{2}F\left(z\right)\nn\\
=&-\frac{4\textrm{i}}{p^{3}}e^{-\mathrm{i}pz_{0}}F\left(z_{0}\right)+\mathcal{O}\left(\frac{1}{p^{4}}\right),\label{eq:asymptotics-2}
\end{align}
where $p=2\pi s/L$ and $s$ is an integer. 
For multiple
singular points, the asymptotic of the Fourier transform of the wave function is  given by the sum of the corresponding  terms  (\ref{eq:asymptotics-0}), (\ref{eq:asymptotics-1}),
and (\ref{eq:asymptotics-2}). 
Using (\ref{eq:wave-function-expansion}), and (\ref{eq:asymptotics-0}),
(\ref{eq:asymptotics-1}), (\ref{eq:asymptotics-2}), the momentum
representation of the wave function with respect to  the first particle reads 
\begin{align}
& \psi\left(p,z_{2},...,z_{N}\right) \nn\\
= & \frac{1}{\sqrt{2\pi}}\int_{0}^{L}dz_{1}e^{-\mathrm{i}pz_{1}}\psi\left(z_{1},z_{2},...,z_{N}\right)\nonumber \\
  \overset{\forall2\leq i\leq N}{=} & \frac{1}{\sqrt{2\pi}}\int_{0}^{L}dz_{1}e^{-\mathrm{i}pz_{1}}\left[\psi_{0}\left(Z_{i}\right)\textrm{sgn}\left(z_{i1}\right)+\psi_{1}\left(Z_{i}\right)z_{i1}+\psi_{2}\left(Z_{i}\right)\textrm{sgn}\left(z_{i1}\right)z_{i1}^{2}+\cdots\right]\nonumber \\
  \overset{\left|p\right|\rightarrow\infty}{=} & \sum_{i=2}^{N}\frac{1}{\sqrt{2\pi}}e^{-\mathrm{i}pz_{i}}\left[\left(\frac{2\mathrm{i}}{p}+\frac{2}{p^{2}}\partial_{z_{1}}-\frac{2\textrm{i}}{p^{3}}\partial_{z_{1}}^{2}\right)\psi_{0}\left(Z_{i}\right)-\frac{4\textrm{i}}{p^{3}}\psi_{2}\left(Z_{i}\right)\right]_{z_{1}=z_{i}-\varepsilon}
\label{eq:phi-p}
\end{align}
which can be used to compute the momentum distribution in an analytical way.

The momentum distribution is obtained by a multiple integral of $\left|\psi\left(p,z_{2},...,z_{N}\right)\right|^{2}$
with respect to $z_{2},z_{3},\cdots,z_{N}$
\begin{eqnarray}
w\left(p\right)&=&N\int_{0}^{L}dz_{2}\cdots\int_{0}^{L}dz_{N}\left|\psi\left(p,z_{2},...,z_{N}\right)\right|^{2}\nn\\
&\overset{\left|p\right|\rightarrow\infty}{=}&\frac{C_{2}}{p^{2}}+\frac{C_{3}}{p^{3}}+\frac{C_{4}}{p^{4}}, \label{eq:w-p}
\end{eqnarray}
where the coefficients $C$'s are regarded  as the Contacts \cite{Tan2008_1,Tan2008_2,Tan2008_3}, namely 
\begin{align}
C_{2} &=\frac{2N\left(N-1\right)}{\pi}\int_{0}^{L}dz_{2}\cdots\int_{0}^{L}dz_{N}\left|\psi_{0}\left(Z_{2}\right)\right|_{z_{1}=z_{2}-\varepsilon}^{2},\label{eq:alpha02}\\
C_{3} &=\frac{4N\left(N-1\right)}{\pi}\int_{0}^{L}dz_{2}\cdots\int_{0}^{L}dz_{N}\textrm{Im}\left(\psi_{0}^{*}\left(Z_{2}\right)\partial_{z_{1}}\psi_{0}\left(Z_{2}\right)\right)|_{z_{1}=z_{2}-\varepsilon},\label{eq:alpha03}\\
C_{4} &=\frac{2N\left(N-1\right)}{\pi}\int_{0}^{L}dz_{2}\cdots\int_{0}^{L}dz_{N}\left.[\left|\partial_{z_{1}}\psi_{0}\left(Z_{2}\right)\right|^{2}
-2\textrm{Re}\left(\psi_{0}^{*}\left(Z_{2}\right)\left[\partial_{z_{1}}^{2}\psi_{0}\left(Z_{2}\right)+2\psi_{2}\left(Z_{2}\right)\right]\right)]\right|_{z_{1}=z_{2}-\varepsilon}.\label{eq:alpha04}
\end{align}

Here $\textrm{Re}\left(x\right)$, $\textrm{Im}\left(x\right)$ denote
the real part and imaginary part of the function $x$, respectively. 

In order to evaluate the large-moment tail of momentum distribution, we define
the two body correlation function
\begin{equation}
g_{2}\left(y_{1},y_{2};z_{1},z_{2}\right)=N\left(N-1\right)\frac{\int_{0}^{L}dz_{3}\cdots\int_{0}^{L}dz_{N}\psi^{*}\left(y_{1},y_{2},z_{3},\cdots,z_{N}\right)\psi\left(z_{1},z_{2},z_{3},\cdots,z_{N}\right)}{\int_{0}^{L}dz_{1}\cdots\int_{0}^{L}dz_{N}\left|\psi\left(z_{1},z_{2},z_{3},\cdots,z_{N}\right)\right|^{2}}. \label{eq:correlation-function}
\end{equation}
Due to the symmetry of the wave function, the local two-body correlation function $g_{2}\left(0,0;0,0\right)$ vanishes.
However, the $p$-wave conditions impose a discontinuity of the  wave function in the vicinity of interaction point. 
Thus  the quasi-local two-body
correlation function $g_{2}\left(0,\varepsilon;0,\varepsilon\right)$ reveals 
the nature of p-wave contacts. 
It gives  the probability of finding two fermions  staying in a short distance $\varepsilon$.
It appears to be  nonzero for finite interaction strength. 
For the  homogeneous system,  the two-body correlation function is translational invariant,
and therefore $C_{2}$ is proportional to the quasi-local two-body correlation function.
Whereas  other contacts $C_{3}$ and $C_{4}=C_{4}^{c}+C_{4}^{r}$ can be expressed in terms of 
derivatives of the two-body correlation function, namely
\begin{align}
C_{2} &=\frac{2L}{\pi}g_{2}\left(0,\varepsilon;0,\varepsilon\right),\label{eq:alpha2}\\
C_{3} &=\frac{2L}{\pi}\left[\textrm{Im}\partial_{z_{2}}\right]g_{2}\left(y_{2},y_{2}+\varepsilon;z_{2},z_{2}+\varepsilon\right)|_{y_{2}=z_{2}=0},\label{eq:alpha3}\\
C_{4}^{c} &=\frac{L}{2\pi}\left[\partial_{y_{2}}\partial_{z_{2}}
-2\textrm{Re}\partial_{z_{2}}^{2}\right]g_{2}\left(y_{2},y_{2}+\varepsilon;z_{2},z_{2}+\varepsilon\right)|_{y_{2}=z_{2}=0},\label{eq:alpha4_c}\\
C_{4}^{r} &=-\frac{L}{\pi}\left[\textrm{Re}\left(\partial_{z_{2}}-\partial_{z_{1}}\right)^{2}\right]g_{2}\left(y_{2},y_{2}+\varepsilon;z_{1},z_{2}+\varepsilon\right)|_{y_{2}=z_{2}=z_{1}=0}.\label{eq:alpha4_r}
\end{align}
Here $C_{4}^{c}$ is related to the derivatives of correlation function with respect to its two
coordinates together, which is related to the center of mass  movements of the pairs. While $C_{4}^{r}$ is related to the difference of the derivatives of correlation function
with respect to $z_{1}$ and $z_{2}$, indicating a contribution from the
relative motion of the pairs. For the ground state or thermodynamical
equilibrium state without breaking inversion symmetry, the momentum distribution is symmetric about $p=0$,
and it is obvious that $C_{3}=0$, which indeed the case after an explicit calculation. Consequently, the momentum distribution
at a large momentum tail  has two terms
\begin{equation}
w\left(p\right)\overset{\left|p\right|\rightarrow\infty}{=}\frac{C_{2}}{p^{2}}+\frac{C_{4}}{p^{4}}, \label{eq:wp-T}
\end{equation}
which determines the two p-wave contacts.

\begin{figure}
\begin{centering}
\includegraphics[scale=0.5]{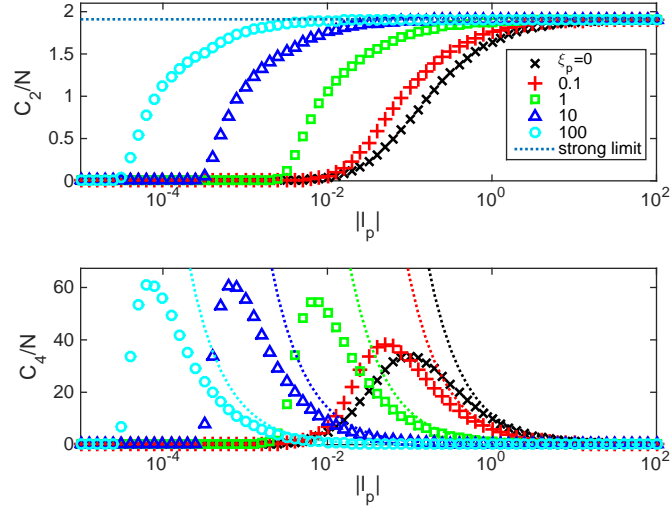}
\par\end{centering}
\caption{\label{fig:four_particles}
The contacts of $C_2$ and $C_{4}$ at 
 the ground state of the model with particle number $N=4$.
 The symbols denote the numerical curves determined  from equations
(\ref{eq:alpha2}), (\ref{eq:alpha4_c}), and (\ref{eq:alpha4_r}).
Wheres the dashed lines are  the strong coupling result from 
(\ref{eq:C2_C4_strong}).  }
\end{figure}

In Figure \ref{fig:four_particles} we show the numerical result of $C_{2}$
and $C_{4}$ for the ground state of the model with four particles. 
When the scattering length  $l_{p}=0$, the model behaves as the ideal Fermi gas with the zero values of  contacts $C_{2}=C_{4}=0$. 
When the scaling length $|l_{p}|$ increases, the fermions prefer to stay together due to the attractive
interaction that leads  to an increase  of the contact  $C_{2}$.
Here we observe that  $C_{2}$ increases
more quickly for a larger value of the range  $\xi_{p}$. 
When $l_{p}\rightarrow\infty$,
$C_{2}$ saturates to the limit of strongly interacting case, which are the same for various $\xi_{p}$'s.
On the other hand, with the increase of $|l_{p}|$, $C_{4}$ first grows to a maximal value and then decreases  rapidly to zero. The peak positions  of $C_{4}$ move to small values of  $|l_{p}|$ for increasing value of effective range $\xi_{p}$.

\section{Correlation Function}
\label{CF}

In this section, we present a straightforward calculation of the contact $C_{2}$, and $C_{4}$ for weak interaction regime, i.e.  $\left|l_{p}\right|\ll1/2\pi n$ and $\xi_{p}<1/2\pi n$, where the particle number density $n=N/L$. 
Following  the method used  for calculating high order local and nonlocal correlation functions of 1D strong interaction Bose gas \cite{correlation_Nandani2016}, we may directly calculate the correlation functions of the 1D  p-wave Fermi gases up to  the second order of  $\left|l_{p}\right|$ .

\subsection{Weak Interaction}

For weak interaction, i.e. $\left|l_{p}\right|\ll1/2\pi n$ and $\xi_{p}<1/2\pi n$, the coefficients $a\left(P\right)$ in the wave function
can be expanded up to the second order of $\left|l_{p}\right|$ with
the following form
\begin{equation}
a\left(P\right)\approx\left(-1\right)^{P}\prod_{i<j=1}^{N}\left(1-\left|l_{p}\right|\textrm{i}\left(\lambda_{P_{i}}-\lambda_{P_{j}}\right)-\frac{1}{2}l_p^2\left(\lambda_{P_{i}}-\lambda_{P_{j}}\right)^{2}\left[1+2\textrm{i}\xi_{p}\left(\lambda_{P_{i}}-\lambda_{P_{j}}\right)\right]\right),\label{eq:AP-l}
\end{equation}
and as a result the wave function in the domain $\left\{ 0\leq z_{1}\leq\cdots\leq z_{N}\leq L\right\} $
for $z_{2}=z_{1}+\varepsilon$ has the new form 
\begin{equation}
\psi\left(z_{1},\cdots,z_{N}\right)=\psi^{\left(1\right)}\left(z_{1},\cdots,z_{N}\right)\left|l_{p}\right|+\psi^{\left(2\right)}\left(z_{1},\cdots,z_{N}\right)l_p^2,\label{eq:phi-l}
\end{equation}
in which the wave function to the first order and the second order of
$\left|l_{p}\right|$ are given by 
\begin{equation}
\psi^{\left(1\right)}\left(z_{1},\cdots,z_{N}\right)=\left(\partial_{z_{2}}-\partial_{z_{1}}\right)\psi^{\left(0\right)}\left(z_{1},\cdots,z_{N}\right),\label{eq:phi-l-1}
\end{equation}
\begin{equation}
\psi^{\left(2\right)}\left(z_{1},\cdots,z_{N}\right)=-\left[\xi_{p}\left(\partial_{z_{2}}-\partial_{z_{1}}\right)^{2}+\left(N-2\right)\left(\partial_{z_{2}}+\partial_{z_{1}}\right)+\sum_{i=3}^{N}\left(N+1-2i\right)\partial_{z_{i}}\right]\psi^{\left(1\right)}\left(z_{1},\cdots,z_{N}\right),\label{eq:phi-l-2}
\end{equation}
respectively. Here the zeroth order wave function $\psi^{\left(0\right)}$ has the form of a Slater determinant $\psi^{\left(0\right)}\left(z_{1},\cdots,z_{N}\right)=\sum_{P}\left(-1\right)^{P}e^{\mathrm{i}\sum_{j=1}^{N}\lambda_{P_{j}}z_{j}}$. Without loss of generality we shall assume that $\lambda_{1}<\lambda_{2}<\cdots<\lambda_{N}$. Using the wave function
(\ref{eq:phi-l}), we can show that  up to the order of $\left|l_{p}\right|^3$, the numerator of the two body correlation function (\ref{eq:correlation-function}) for a positive
infinitesimal $\varepsilon$ reads (for details, see calculation in Appendix A).
\begin{equation}
g_{2}\left(y_{1},y_{2};z_{1},z_{2}\right)=\left\{ 1-\left|l_{p}\right|\xi_{p}\left[\left(\partial_{y_{2}}-\partial_{y_{1}}\right)^{2}+\left(\partial_{z_{2}}-\partial_{z_{1}}\right)^{2}\right]\right\} g_{2}^{\left(2\right)}\left(y_{1},y_{2};z_{1},z_{2}\right).\label{eq:g2-l}
\end{equation}
Here the second order correlation function is defined as the following derivatives 
\begin{equation}
g_{2}^{\left(2\right)}\left(y_{1},y_{2};z_{1},z_{2}\right)=l_p^2\left(\partial_{y_{2}}-\partial_{y_{1}}\right)\left(\partial_{z_{2}}-\partial_{z_{1}}\right)g_{2}^{\left(0\right)}\left(y_{1},y_{2};z_{1},z_{2}\right).\label{eq:g2-2-l}
\end{equation}
In the above calculation, we have omitted  higher order of $|l_p| \xi_p$ terms. 
We define the zeroth order correlation function
\begin{equation}
g_{2}^{\left(0\right)}\left(y_{1},y_{2};z_{1},z_{2}\right)=\int_{0}^{L}dz_{3}\cdots\int_{0}^{L}dz_{N}\left[\psi^{\left(0\right)}\left(y_{1},y_{2},z_{3},\cdots,z_{N}\right)\right]^{*}\psi^{\left(0\right)}\left(z_{1},z_{2},z_{3},\cdots,z_{N}\right).\label{eq:g2-0}
\end{equation}
This zeroth order correlation function  goes to zero when we set up the short distance limit $\varepsilon \rightarrow 0$ (due to  the zeroth order wave function $\psi^{\left(0\right)}\left(z_{1},\cdots,z_{N}\right)$). 
However, its derivatives gives non-zero results when taking the limit $\varepsilon\to 0$.

 From BA equations (\ref{eq:BA-equations}) with the weak interaction,
we can obtain the solution $\lambda_{i}=\lambda_{i}^{F}\alpha$ for
total momentum $\sum_{i}\lambda_{i}=0$, where $\lambda_{i}^{F}=2\pi J_{i}/L$,
$\alpha=1-2\left|l_{p}\right|n$, and the $J_{i}$ are (half-)integers
satisfying $J_{1}<J_{2}<...<J_{N}$. Under the scaling $z_{i}=z_{i}^{F}/\alpha$, the zeroth order wave function $\psi^{\left(0\right)}\left(z_{1},\cdots,z_{N}\right)$
is identical to the form for ideal Fermi gas $\psi^{F}\left(z_{1}^{F},\cdots,z_{N}^{F}\right)=\sum_{P}\left(-1\right)^{P}e^{\mathrm{i}\sum_{j=1}^{N}\lambda_{P_{j}}^{F}z_{j}^{F}}$.
We also find  the normalization condition \cite{correlation_Nandani2016}
\begin{equation}
\int_{0}^{L}dz_{1}\cdots\int_{0}^{L}dz_{N}\left|\psi\left(z_{1},\cdots,z_{N}\right)\right|^{2}=\alpha^{1-N}
\int_{0}^{L}dz_{1}^{F}...\int_{0}^{L}dz_{N}^{F}\left|\psi^F\left(z_{1}^{F},\cdots,z_{N}^{F}\right)\right|^{2}.\label{eq:normalization}
\end{equation}
Then the zeroth order correlation function with normalization factor has
the following form
\begin{equation}
g_{2}^{\left(0\right)}\left(y_{1},y_{2};z_{1},z_{2}\right)=\alpha N\left(N-1\right)\frac{\int_{0}^{L}dz_{3}^{F}\cdots\int_{0}^{L}dz_{N}^{F}\left[\psi^{F}\left(y_{1}^{F},y_{2}^{F},z_{3}^{F},\cdots,z_{N}^{F}\right)\right]^{*}\psi^{F}\left(z_{1}^{F},z_{2}^{F},z_{3}^{F},\cdots,z_{N}^{F}\right)}{\int_{0}^{L}dz_{1}^{F}...\int_{0}^{L}dz_{N}^{F}\left|\psi^{F}\left(z_{1}^{F},\cdots,z_{N}^{F}\right)\right|^{2}}.\label{eq:g2-0-Fermi2}
\end{equation}
Since $\psi^{F}\left(z_{1}^{F},\cdots,z_{N}^{F}\right)$
is a Slater determinant, one can use  Wick's theorem \cite{correlation_Nandani2016}
\begin{equation}
g_{2}^{\left(0\right)}\left(y_{1},y_{2};z_{1},z_{2}\right)=\alpha\left[G\left(y_{1}^{F},z_{1}^{F}\right)G\left(y_{2}^{F},z_{2}^{F}\right)-G\left(y_{1}^{F},z_{2}^{F}\right)G\left(y_{2}^{F},z_{1}^{F}\right)\right],\label{eq:g2-0-Green}
\end{equation}
where the single particle reduced density matrix of ideal fermions is given by $G\left(y,z\right)=\frac{1}{L}\sum_{i=1}^{N}\exp\left[-\textrm{i}\lambda_{i}^{F}\left(y-z\right)\right]$.
Substituting the above formula into expression (\ref{eq:g2-2-l}),
then substituting (\ref{eq:g2-2-l}) into equation (\ref{eq:g2-l}) and after a lengthy calculation, 
we obtain the two body correlation function (for details, see calculation in Appendix A)
\begin{equation}
g_{2}\left(y_{1},y_{2};z_{1},z_{2}\right)=l_p^2\frac{1}{L^{2}}\sum_{i,j=1}^{N}\left[\alpha^{3}\left(\lambda_{i}^{F}-\lambda_{j}^{F}\right)^{2}+2\left|l_{p}\right|\alpha^{5}\xi_{p}\left(\lambda_{i}^{F}-\lambda_{j}^{F}\right)^{4}\right]X\left(\lambda_{i}^{F},\lambda_{j}^{F}\right),\label{eq:g2-Green}
\end{equation}
where  $X\left(\lambda_{i}^{F},\lambda_{j}^{F}\right)=\exp\left[-\textrm{i}\lambda_{i}^{F}\left(y_{1}^{F}-z_{1}^{F}\right)\right]\exp\left[-\textrm{i}\lambda_{j}^{F}\left(y_{2}^{F}-z_{2}^{F}\right)\right]+\exp\left[-\textrm{i}\lambda_{i}^{F}\left(y_{1}^{F}-z_{2}^{F}\right)\right]\exp\left[-\textrm{i}\lambda_{j}^{F}\left(y_{2}^{F}-z_{1}^{F}\right)\right]$.
We will use this expression to  calculate the p-wave contacts. 

\begin{figure}
\begin{centering}
\includegraphics[scale=0.5]{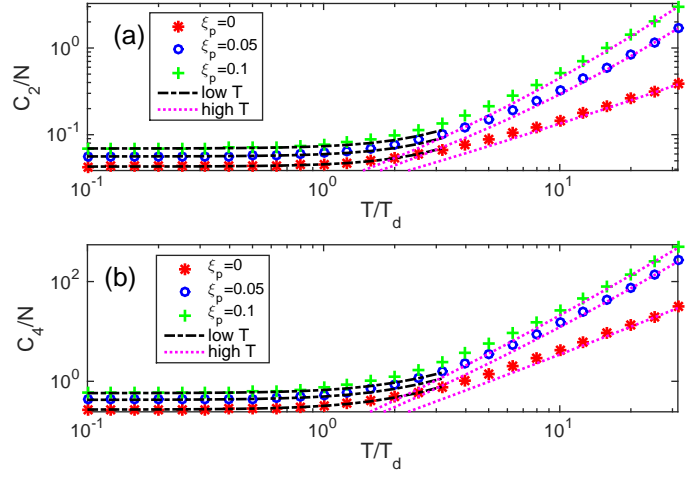}
\par\end{centering}
\caption{The Contacts for p-wave Fermi gas with weakly attractive interaction
in the thermodynamic limit for $\left\vert l_{p}\right\vert =0.1$ and $n=1$. The marked-lines
are exact results (\ref{eq:alpha2-T}) and (\ref{eq:alpha4-T}).
The black dash-dot-lines are for low temperature limit
(\ref{eq:alpha2-lowT}) and (\ref{eq:alpha4-lowT}). The magenta dot-lines
are for high temperature limit (\ref{eq:alpha2-highT})
and (\ref{eq:alpha4-highT}). The Contacts $C_{2}$ and $C_{4}$ increase with both the temperature and
the effective range.}
\label{Figure2}
\end{figure}

\subsection{P-wave contacts}

In the thermodynamic limit, we may use
the Fermi distribution function to evaluate the p-wave contact through the relations given in (\ref{eq:alpha2})-(\ref{eq:alpha4_r}). 
The modified Fermi distribution function with single particle energy $\left(\lambda^{F}\alpha\right)^{2}$ is given by 
\begin{equation}
f\left(\lambda^{F}\right)=\frac{1}{1+\exp\left(\left[\left(\lambda^{F}\alpha\right)^{2}-A\right]/T\right)}, 
\end{equation}
here $T$ and $A$ are  the temperature and the effective chemical potential of the 1D
Fermi gas respectively, and we have set Boltzmann constant $k_{B}=1$.
It satisfies the normalization condition $\int_{-\infty}^{\infty}f\left(\lambda^{F}\right)\frac{L}{2\pi}d\lambda^{F}=N$, which include the first order modification of $\left\vert l_{p}\right\vert$ and is equivalent to the corresponding result from thermodynamical Bethe ansatz equations, see Appendix C.
Notice that  the density of states, i.e. the number of particles with fermionic momenta in the interval 
$(\lambda^F,\lambda^{F}+d\lambda^{F})$, is $(L/2 \pi)f\left(\lambda^{F}\right)d \lambda^{F}$, such that the summation in the two-body 
correlation function (\ref{eq:g2-Green}) becomes an integral in the following form 
\begin{equation}
g_{2}\left(y_{1},y_{2};z_{1},z_{2}\right)=\frac{l_p^2}{4\pi^{2}}\int_{-\infty}^{\infty}d\lambda_{1}^{F}\int_{-\infty}^{\infty}d\lambda_{2}^{F}f\left(\lambda_{1}^{F}\right)f\left(\lambda_{2}^{F}\right)\left[\alpha^{3}\left(\lambda_{1}^{F}-\lambda_{2}^{F}\right)^{2}+2\left|l_{p}\right|\alpha^{5}\xi_{p}\left(\lambda_{1}^{F}-\lambda_{2}^{F}\right)^{4}\right]X\left(\lambda_{1}^{F},\lambda_{2}^{F}\right).\label{eq:g2-T}
\end{equation}
For our convenience in calculation, we make a change of variable $\lambda_{1,2}^{F}=2\pi nx_{1,2}$ and then  we obtain 
\begin{eqnarray}
g_{2}\left(y_{1},y_{2};z_{1},z_{2}\right)&=&l_p^24\pi^{2}n^{4}\int_{-\infty}^{\infty}dx_{1}\int_{-\infty}^{\infty}dx_{2}\mathscr{\mathcal{N}}\left(x_{1}\right)\mathcal{N}\left(x_{2}\right)\left[\alpha^{3}\left(x_{1}-x_{2}\right)^{2}+2\left|l_{p}\right|\alpha^{5}\xi_{p}\left(2\pi n\right)^{2}\left(x_{1}-x_{2}\right)^{4}\right]\nonumber\\
&& \times X\left(2\pi nx_{1},2\pi nx_{2}\right),\label{eq:g2-Tx}
\end{eqnarray}
where the function  $\mathcal{N}\left(x\right)=f\left(2\pi nx\right)$  is subject to the
normalization condition $\int_{-\infty}^{\infty}\mathcal{N}\left(x\right)dx=1$. 
Substituting the above formula into equations (\ref{eq:alpha2}-\ref{eq:alpha4_r}), we obtain explicitly  the p-wave contacts in terms of the scattering length  $\left\vert l_p \right\vert$ and the effective range $\xi_p$
\begin{eqnarray}
C_{2}&=&32\pi Nn^{3}l_p^2\left[\left(1-6\left|l_{p}\right| n\right)f_{2}+2\left|l_{p}\right|\xi_{p}\left(2\pi n\right)^{2}\left(f_{4}+3f_{2}^{2}\right)\right],\label{eq:alpha2-T}\\
C_{4}&=&32\pi^{3}Nn^{5}l_p^2\left[\left(1-10\left|l_{p}\right| n\right)\left(5f_{4}+3f_{2}^{2}\right)+2\left|l_{p}\right|\xi_{p}\left(2\pi n\right)^{2}\left(5f_{6}+27f_{2}f_{4}\right)\right],\label{eq:alpha4-T}
\end{eqnarray}
with $f_{i}=\int_{-\infty}^{\infty}dx\mathcal{N}\left(x\right)x^{i}=\frac{1}{\left(2\pi n\alpha\right)^{i+1}}\int_{0}^{\infty}dy\frac{y^{\left(i-1\right)/2}}{1+e^{\left(y-A\right)/T}}$.
We observe that  that in the weak coupling regime the p-wave contacts $C_{2}$ and $C_{4}$ increase with both the scattering length $\left|l_{p}\right|$ and the effective range $\xi_{p}$. 
In addition we have checked that indeed $C_{3}=0$ for the ground state and  equilibrium states. 

At low temperature $T\ll T_{d}$, here  $T_{d}=\frac{\hbar^{2}n^{2}}{2mk_{B}}$ is the degeneracy temperature, we can further calculate the contacts by the Sommerfeld expansion
\begin{equation}
C_{2}=\frac{8}{3}\pi Nn^{3}l_p^2\left[\left(1-6\left|l_{p}\right| n\right)+\left(1+2\left|l_{p}\right| n\right)\frac{1}{4\pi^{2}}\tau^{2}+\frac{16}{5}\left|l_{p}\right|\xi_{p}\left(\pi n\right)^{2}\left(1+\frac{5}{8\pi^{2}}\tau^{2}\right)\right],\label{eq:alpha2-lowT}
\end{equation}
 \begin{equation}
C_{4}=\frac{8}{3}\pi^{3}Nn^{5}l_p^2\left[\left(1-10\left|l_{p}\right| n\right)+\left(1-2\left|l_{p}\right| n\right)\frac{3}{4\pi^{2}}\tau^{2}+\frac{132}{35}\left|l_{p}\right|\xi_{p}\left(\pi n\right)^{2}\left(1+\frac{14}{11\pi^{2}}\tau^{2}\right)\right],\label{eq:alpha4-lowT}
\end{equation}
with $\tau=T/T_{d}$, see Appendix B. 
This indicates a simple relationship between the leading terms of $C_{2}$ and $C_{4}$, i.e.  $C_{4}=C_{2}\left(n\pi\right)^{2}$.
In this weak coupling regime, the contact $C_2$ of the spinless p-wave  Fermi gas behaves much likes the one of the Lieb-Liniger Bose gas with a strong repulsion \cite{correlation_Nandani2016}.
The contacts  $C_{2}$ and $C_{4}$ increase smoothly as we increase the temperature. 
At relative high temperatures $T\gg T_{d}$, one can apply  the Taylor series expansion with the distribution  function $f_{i}$ under the condition  $\exp\left(A/T\right)\ll1$,
and then one has 
\begin{eqnarray}
C_{2}&=&\frac{4}{\pi}\tau Nn^{3}l_p^2\left[\left(1-2\left|l_{p}\right| n\right)+\left(1-4\left|l_{p}\right| n\right)\frac{\sqrt{2\pi}}{2}\tau^{-1/2}+6\left|l_{p}\right|\xi_{p}n^{2}\tau\left(1+\frac{7\sqrt{2\pi}}{8}\tau^{-1/2}\right)\right],\label{eq:alpha2-highT}\\
C_{4}&=&\frac{9}{\pi}\tau^{2}Nn^{5}l_p^2\left[\left(1-2\left|l_{p}\right| n\right)+\left(1-4\left|l_{p}\right| n\right)\frac{19\sqrt{2\pi}}{24}\tau^{-1/2}+\frac{26}{3}\left|l_{p}\right|\xi_{p}n^{2}\tau\left(1+\frac{445\sqrt{2\pi}}{416}\tau^{-1/2}\right)\right].\label{eq:alpha4-highT}
\end{eqnarray}
In the above calculation, the  weak interaction conditions require that $\left|l_{p}\right|/\Lambda\ll1$, where 
the thermal de Broglie wavelength $\Lambda=\sqrt{\frac{4\pi}{k_{B}T}}$. 
Namely, we request  $T\ll T_{d}/\left(l_{p}n\right)^{2}$ and $C_{2}/\left(Nn\right)\ll1$.
 We show these low and 
high temperature behaviour of the contacts of the spinless Fermi gases with an attractive p-wave interaction  in Figure \ref{Figure2}, 
where a good agreement between the numerical result and  the exact result which we have obtained here.

\subsection{Energy derivatives with respect to the scattering length}
To related the $p$-wave contact defined using asymptotic behavior of the momentum distribution to other physical observables, let us consider how $C_2$ is related to the derivative of energy with respect to $l_p$. Since the sub-leading $C_4$ term involves two contacts $C_4^r$ and $C_4^c$, and only the former is related to energy derivative with respect to the effective range, we shall not discuss that adiabatic relation in this work. To that end, let us consider the $p$-wave pseudopotential of the following form~\cite{Psuedopotential_p_wave_Girardeau2004, Pseudopotential_p_wave_Grosse2004} 
\begin{equation}
V\left(z_{1},\cdots z_{N}\right)=-a_{p}^{1D}\sum_{i<j}\left(\frac{\partial}{\partial z_{i}}-\frac{\partial}{\partial z_{j}}\right)\delta\left(z_{i}-z_{j}\right)\left(\frac{\partial}{\partial z_{i}}-\frac{\partial}{\partial z_{j}}\right).\label{eq:potential_weak}
\end{equation}


According to the Hellman-Feynman theorem, the two body correlation function
is related to derivative of energy with respect to $a_{p}^{\textrm{1D}}$, i.e.
$g_{2}\left(0,\varepsilon;0,\varepsilon\right)=\frac{\left(a_{p}^{\textrm{1D}}\right)^{2}}{2L}\frac{\partial E}{\partial a_{p}^{\textrm{1D}}}$.
Combining the definition of contact $C_{2}$ with the effective scattering length $l_{p}=a_{p}^{\mathrm{1D}}/2$, we may verify that the
following relationship holds:  
$C_{2}=\frac{1}{\pi}\frac{\partial E}{\partial\left(-a_{p}^{1D}\right)^{-1}}$.
For a grand canonical ensemble, this relation can be rewritten as 
\begin{equation}
C_{2}=\frac{L}{\pi}\left.\frac{\partial P}{\partial\left(a_{p}^{1D}\right)^{-1}}\right|_{\mu,T}, \label{eq:C2_E}
\end{equation}
where the pressure  $P$ at finite temperatures and arbitrary an interaction strength can
be obtained by thermodynamical Bethe ansatz (TBA) equations for the model, see Appendix C. 
Then  one can recover the two-body correlation function, and thereafter the Contact $C_{2}$ given in (\ref{eq:alpha2-lowT}) and (\ref{eq:alpha2-highT}).

 Here we firstly study  the ground energy and its  corresponding contact using the BA equations,
then further  analytically calculate  the finite temperature  contact using the  TBA equations.
 For the ground state under  condition $\left|l_p\right|\ll L$ and $\xi_p\ll L$, the BA equations (\ref{eq:BA-equations2})  give a class of  asymptotic solutions 
 \begin{equation}
 \lambda_{i}=\lambda_{i}^{F}\left[1-2n\left|l_p\right|+(4n^2-2\xi_{p}(n\left(\lambda_{i}^{F}\right)^{2}+3E^{F}/L))l_{p}^{2}\right]
 \end{equation}
  up to the  second order of $\left|l_{p}\right|$, 
  where $\lambda_{i}^{F}$ and $E^{F}$ are quasi-momentum and energy for 1D ideal Fermi gases, respectively. 
  Consequently, we have  the ground state  energy 
  \begin{equation}
  E=E^{F}\left[1-4n\left|l_p\right|+4(3n^2-\xi_{p}(n\sum_{i=1}^{N}\left(\lambda_{i}^{F}\right)^{4}/E^{F}+3E^{F}/L))l_{p}^{2}\right].
  \end{equation}
 Taking derivative of the ground state energy with respect to 1D scattering length, we  obtain the contact 
 \begin{equation}
 C_{2}=\frac{1}{\pi}8nE^{F}l_{p}^{2}\left[1-6n\left|l_p\right|+2\xi_{p}\left|l_p\right|(\sum_{i=1}^{N}\left(\lambda_{i}^{F}\right)^{4}/E^{F}+3E^{F}/N))\right].
 \end{equation}
  In the thermodynamics limit, the contact $C_{2}=\frac{8}{3}\pi n^{3}N l_{p}^{2}\left(1-6n\left|l_p\right|+\frac{16}{5}\xi_{p}\left|l_p\right|\pi^2 n^2\right)$ is consistent with expression given in (\ref{eq:alpha2-lowT}) for the ground state.

In the following we consider the grand canonical ensemble in order to confirm the relation (\ref{eq:C2_E}).
For $l_{p}<0$, all solutions of the BA equations  Eq. (\ref{eq:BA-equations}) are real.
However, at finite temperatures, the eigenstates become degenerate. 
Following the approach to the thermodynamics   of  the 1D Bose gas introduced by C. N. Yang and C. P. Yang \cite{Yang1969},
here the Gibbs free energy  $G=E-TS-\mu N$  of the spinless Fermi gases with an attractive p-wave interaction  can be  determined   by a minimization the Gibbs energy, i.e.  $\delta G$ with respect to the BA root densities.
In the Gibbs free energy, $S$ denotes  entropy and $\mu$ is the chemical potential.
It follows that 
\begin{equation}
\epsilon(\lambda)=\lambda^{2}-\mu-\frac{T}{2\pi}\int_{-\infty}^{\infty}K\left(k-\lambda\right)\mathrm{ln}\left\{ 1+\mathrm{exp}\left[-\epsilon\left(k\right)/T\right]\right\} dk,\label{eq:TBA}
\end{equation}
where $\epsilon(\lambda)$ is the dressed energy characterizing the excitation energy. 
In the  above equation, the kernel function $K(x)$ is given by 
\begin{equation}
K(x)=\frac{2|l_{p}|(|l_{p}|\xi_{p}x^{2}+1)}{(|l_{p}|\xi_{p}x^{2}-1)^{2}+l_{p}^{2}x^{2}}.
\end{equation}
The pressure $P$ can then be expressed as
\begin{equation}
P=\frac{T}{2\pi}\int_{-\infty}^{\infty}\mathrm{ln}\left\{ 1+\mathrm{exp}\left[-\epsilon\left(\lambda\right)/T\right]\right\} d\lambda. \label{eq:pressure}
\end{equation}
This gives the equation of states, from which  the particle density and the compressibility are  determined by $n=\frac{\partial P}{\partial\mu}$ and $\kappa=\frac{\partial n}{\partial\mu}$, respectively. 

In  the weak interaction limit ($\left|l_{p}\right|\ll1$), one can expand the kernel
function $K(x)$ in powers of scattering length $\left|l_{p}\right|$. Up to the first few leading terms, the TBA equation (\ref{eq:TBA})  can be simplified as
\begin{equation}
\epsilon(\lambda)=B\lambda^{2}-A\label{eq:TBA_lp}
\end{equation}
where the effective chemical potential  $A=\mu+2Q_{0}\left|l_{p}\right|+6\xi_{p}Q_{2}l_{p}^{2}$ with the  notations $B=1-6\xi_{p}Q_{0}l_{p}^{2}$
and $Q_{j}=\frac{T}{2\pi}\int_{-\infty}^{\infty}k^{j}\mathrm{ln}\left\{ 1+\mathrm{exp}\left[-\frac{\epsilon\left(k\right)}{T}\right]\right\} dk$.

In the low temperature limit ($T\ll T_{d}$), the pressure $P$ up to the
second order of $\left|l_{p}\right|$ reads
\begin{equation}
P=\frac{2}{3\pi}\mu^{3/2}\left\{ 1+2\frac{\sqrt{\mu}}{\pi}\left|l_{p}\right|+\left(\frac{14}{3}+\frac{16\pi}{5}\sqrt{\mu}\xi_{p}\right)\frac{\mu}{\pi^{2}}l_{p}^{2}+\frac{\pi^{2}}{8}\frac{T^{2}}{\mu^{2}}\left[1+\frac{4}{3}\frac{\sqrt{\mu}}{\pi}\left|l_{p}\right|+\left(\frac{10}{3}+\frac{48\pi}{5}\sqrt{\mu}\xi_{p}\right)\frac{\mu}{\pi^{2}}l_{p}^{2}\right]\right\}.\label{eq:P_lowT}
\end{equation}
This gives an important indication of the deviation from the free fermions. 
From the  formula (\ref{eq:C2_E}), the Contact is calculated in a straightforward way 
\begin{equation}
C_2=\frac{8Ll_{p}^{2}}{3\pi^{3}}\mu^{2}\left\{ 1+\left(\frac{14}{3}+\frac{16\pi}{5}\sqrt{\mu}\xi_{p}\right)\frac{\sqrt{\mu}}{\pi}\left|l_{p}\right|+\frac{\pi^{2}}{12}\frac{T^{2}}{\mu^{2}}\left[1+\left(5+\frac{72\pi}{5}\sqrt{\mu}\xi_{p}\right)\frac{\sqrt{\mu}}{\pi}\left|l_{p}\right|\right]\right\}.  \label{eq:g2_lowT}
\end{equation}
In this expression,  the chemical potential can be expressed in terms of the particle density 
\begin{equation}
\mu=\pi^{2}n^{2}\left\{ 1-\frac{16}{3}n\left|l_{p}\right|+4n^{2}l_{p}^{2}\left(5-\frac{16\pi^{2}}{5}n\xi_{p}\right)+\frac{1}{12}\frac{T^{2}}{\pi^{2}n^{4}}\left[1-4n^{2}l_{p}^{2}\left(1+8\pi^{2}n\xi_{p}\right)\right]\right\} .
\end{equation}
The above $p$-wave contact $C_2$  (\ref{eq:g2_lowT}) is indeed consistent with expression
(\ref{eq:alpha2-lowT}), which was obtained from the many-body wave function.
This confirms the universal relation (\ref{eq:C2_E}). 
Furthermore the compressibility has the analytical form
\begin{equation}
\kappa=\frac{1}{2n\pi^{2}}\left(1+\frac{1}{12\pi^{2}}\frac{T^{2}}{n^{4}}\right)\left\{ 1+\left(8+\frac{2}{3\pi^{2}}\frac{T^{2}}{n^{4}}\right)n\left|l_{p}\right|+\left[24+\frac{22}{3\pi^{2}}\frac{T^{2}}{n^{4}}+\left(32\pi^{2}+4\frac{T^{2}}{n^{4}}\right)n\xi_{p}\right]n^{2}l_{p}^{2}\right\}. 
\end{equation}

Furthermore, for a relative high temperature regime, i.e. $T_{d}\ll T\ll T_{d}/\left(l_{p}n\right)^{2}$, up to the second order of $\left|l_{p}\right|$,
the pressure  is given by 
\begin{equation}
P=\frac{1}{2}\frac{ZT^{3/2}}{\sqrt{\pi}}\left\{ 1-\frac{\sqrt{2}}{4}Z+\frac{Z\sqrt{T}}{\sqrt{\pi}}\left|l_{p}\right|\left(1-\frac{3\sqrt{2}}{4}Z\right)+\frac{3}{2\pi}ZTl_{p}^{2}\left[Z-\frac{4\sqrt{2}}{3}Z^{2}+2\sqrt{\pi T}\xi_{p}\left(1-\frac{9\sqrt{2}}{16}Z\right)\right]\right\},\label{eq:P_highT}
\end{equation}
here  fugacity $Z=\exp\left(\mu/T\right)$. 
From the relation  (\ref{eq:C2_E}),
the two-body correlation function is given explicitly 
\begin{equation}
C_2=\frac{T^{2}Z^{2}l_{p}^{2}}{2\pi}\left\{ 1-\frac{3\sqrt{2}}{4}Z+3\sqrt{T/\pi}\left|l_{p}\right|\left[Z-\frac{4\sqrt{2}}{3}Z^{2}+2\sqrt{\pi T}\xi_{p}\left(1-\frac{9\sqrt{2}}{16}Z\right)\right]\right\} .\label{eq:g2_highT}
\end{equation}
It is obvious that this result is consistent with expression (\ref{eq:alpha2-highT}) obtained from the many-body wave function (for details, see calculation in Appendix C). 
This alternative procedure provides a confirmation  of the contact calculated by the many-body wave function  in terms of  the leading and sub-leading terms of the scattering length $\left\vert l_p \right\vert$, see Fig.~\ref{Figure2}.

\subsection{ Strong Interaction Limit}

The case $l_{p}\rightarrow-\infty$ is known as the Fermionic Tonks-Girardeau
gas \cite{Fermi_TG_Bender2005, Fermi_TG_Girardeau2006} at the p-wave resonance.
When, in addition, the effective range tends to zero, the  wave function can be constructed from 
the noninteracting Bose gas with a sign function \cite{Fermi_TG_Bender2005, Fermi_TG_Girardeau2006}. 
In the strong attractive coupling limit, i.e. 
$\left|l_{p}\right|\gg L+2N\xi_p$,  the system has the similar solution
of  the  quasi-momenta for the 1D weakly interacting Bose gas, see review~\cite{Jiang:2015}.
In this case, we assume that $\left|-\xi_{p}\lambda^{2}+1/\left|l_{p}\right|\right|\ll\left|\lambda\right|$ and  find that the phase shift $\theta\left(\lambda\right)$ in the  BA
equations (\ref{eq:BA-equations}) is close to $\pm\pi$, and thus the quasi-momenta $\lambda_{i}$
are proportional to the square root of $1/\left|l_{p}\right|$ in
the form $\lambda_{i}=\beta_{i}H\left(\xi_{p}\right)/\sqrt{\left|l_{p}\right|}$
with $i=1,\cdots,N$, in which the function   $H\left(\xi_{p}\right)=\sqrt{2/\left(L+2N\xi_{p}\right)}$. 
Here  $\left\{ \beta_{i}\right\} $ with $N=1,\,2,\ldots,\,N$  are determined by the sets of equations
$\beta_{i}=\sum_{j=1\left(j\neq i\right)}^{N}1/\left(\beta_{i}-\beta_{j}\right)$,
 which are equivalent to  the roots of the Hermite polynomial of degree $N$.
The summation of  all squares of $\beta_{i}$ gives  $\sum_{i}^{N}\beta_{i}^{2}=N\left(N-1\right)/2$,
so that  the energy of system is given by  $E=N\left(N-1\right)/\left[\left(L+2N\xi_{p}\right)\left|l_{p}\right|\right]$.
It is obvious that  the energy decreases with an  increase of the scattering length $\left|l_{p}\right|$ and the effective range $\xi_{p}$.

The coefficients in the wave function can be rewritten as $a\left(P\right)=\exp\left(\textrm{i}B\left(P\right)/\sqrt{\left|l_{p}\right|}\right)$
with $B\left(P\right)=\sum_{i<j}F\left(\xi_{p},\beta_{P_{i}}-\beta_{P_{j}}\right)$
and $F\left(\xi_{p},\beta\right)=\left(1-\xi_{p}H^{2}\left(\xi_{p}\right)\beta^{2}\right)/\left(H\left(\xi_{p}\right)\beta\right)$.
The wave function in the domain $z_{1}\leq z_{2}\leq...\leq z_{N}$
is reduced into the following form
\begin{equation}
\psi\left(z_{1},z_{2},...,z_{N}\right)=N!-\frac{1}{2}\sum_{P}\left[B\left(P\right)+H\left(\xi_{p}\right)\sum_{j=1}^{N}\beta_{P_{j}}z_{j}\right]^{2}\frac{1}{\left|l_{p}\right|}+\mathcal{O}\left(\left|l_{p}\right|^{-3/2}\right).\label{eq:wavefunction_strong}
\end{equation}
By calculating the quasi-local correlation function (see Appendix D), one obtains
the leading terms for $C_{2}$ and $C_{4}$ as following
\begin{equation}
C_{2}=\frac{2}{\pi}n\left(N-1\right),\qquad C_{4}=\frac{n}{\left|l_{p}\right|}\frac{4\left(N-1\right)^{2}}{\pi L\left(1+2n\xi_{p}\right)}.\label{eq:C2_C4_strong}
\end{equation}
From above expression, we see that  the contact per particle $C_{2}/N$ is a finite value that  
depends only on the particle density in the thermodynamical limit. 
However,
$C_{4}$ decreases with an increase of the scattering length $\left|l_{p}\right|$, and it  goes to zero at the p-wave resonance $\left|l_{p}\right|\rightarrow\infty$.
For $\xi_{p}=0$, $C_{2}$ still has the same result as (\ref{eq:C2_C4_strong}) which
can be obtained from the energy derivative with respect to $\left\vert l_p \right\vert^{-1}$  \cite{Fermi_TG_Bender2005, Fermi_TG_Girardeau2006, contact_p_wave_Sekino2017}. 

\section{Conclusion}

In summary, using the exact man-body  Bethe ansatz wave function, we have  obtained  the large-momentum  tails of momentum distribution of a spinless Fermi gas interacting via $p$-wave scattering. We have found the leading and sub-leading  coefficients of $1/p^{2}$ and $1/p^{4}$, i.e. the $p$-wave contacts, and give their analytic expressions in both weak and strong interaction limit. We show also by explicit calculation how the $p$-wave contacts are related to the short-distance behavior of the two-body correlation functions. In addition, we have obtained the energetics and contacts in both the low and high temperature limits via thermodynamic Bethe ansatz. Our results provide a deep  insight into the nature of the p-wave contacts in one and higher dimensions.  

\acknowledgments
This work has been supported by NSF of China under Grant No. 11704233, No. 11474189 and No. 11674201.
This work is also  supported by Key NNSFC grant number 11534014, the National Key R\&D Program of China  No. 2017YFA0304500. S.Z. is supported by the  AFOSR, ARO, and NSERC. S.Z. is supported by Hong Kong Research Grants Council, Collaborative Research Fund (Grant No. C6026-16W) and GRF 17318316, and the Croucher Foundation under the Croucher Innovation Award. HBS thanks Professor Vladimir Korepin for stimulating discussions, and the support from Society of Interdisciplinary Research (SOIREE).

 \appendix{\bf Appendix}
 \section{Correlation Function}
 In the domain $z_{1}\leq z_{2}\leq...\leq z_{N}$, the wave function
has the following form
\begin{equation}
\psi\left(z_{1},\cdots,z_{N}\right)=\sum_{P}a\left(P\right)e^{\mathrm{i}\sum_{j=1}^{N}\lambda_{P_{j}}z_{j}}.
\end{equation}
For weak interaction, the coefficients $a\left(P\right)$ in above
wave function can be expanded up to the second order of $\left|l_{p}\right|$
\begin{equation}
a\left(P\right)\approx\left(-1\right)^{P}\prod_{i<j}\left(1-\left|l_{p}\right|\textrm{i}\left(\lambda_{P_{i}}-\lambda_{P_{j}}\right)-\frac{1}{2}l_{p}^{2}\left(\lambda_{P_{i}}-\lambda_{P_{j}}\right)^{2}\left[1+2\textrm{i}\xi_{p}\left(\lambda_{P_{i}}-\lambda_{P_{j}}\right)\right]\right).
\end{equation}
When $z_{2}=z_{1}+\varepsilon$, the coefficients $a\left(P_{1}P_{2}P_{3}...P_{N}\right)$
and $a\left(P_{2}P_{1}P_{3}...P_{N}\right)$ share the same plane
wave $e^{\mathrm{i}\left(\lambda_{P_{1}}+\lambda_{P_{2}}\right)z_{1}+\mathrm{i}\sum_{j=3}^{N}\lambda_{P_{j}}z_{j}}$,
so we calculate the summation of them
\begin{align}
 & a\left(P_{1}P_{2}P_{3}...P_{N}\right)+a\left(P_{2}P_{1}P_{3}...P_{N}\right)\nn\\
 & \approx2\left(-1\right)^{P}\left(\frac{\left|l_{p}\right|}{\textrm{i}}\lambda_{P_{12}}+\frac{l_{p}^{2}\xi_{p}}{\textrm{i}}\lambda_{P_{12}}^{3}\right)\prod_{i<j=3}\left(1+\frac{\left|l_{p}\right|}{\textrm{i}}\lambda_{P_{ij}}\right)\nn\\
 & \approx2\left(-1\right)^{P}\left(\frac{\left|l_{p}\right|}{\textrm{i}}\lambda_{P_{12}}+\frac{l_{p}^{2}\xi_{p}}{\textrm{i}}\lambda_{P_{12}}^{3}-l_{p}^{2}\lambda_{P_{12}}\sum_{i<j=3}\lambda_{P_{ij}}\right)\nn\\
 & =2\left(-1\right)^{P}\lambda_{P_{12}}\left(\frac{\left|l_{p}\right|}{\textrm{i}}+l_{p}^{2}\left[\frac{\xi_{p}}{\textrm{i}}\lambda_{P_{12}}^{2}-\left(N-2\right)\left(\lambda_{P_{1}}+\lambda_{P_{2}}\right)-\sum_{i=3}^{N}\left(N+1-2i\right)\lambda_{P_{i}}\right]\right),
\end{align}
in which $\lambda_{P_{ij}}=\lambda_{P_{i}}-\lambda_{P_{j}}$ and $\sum_{i<j=3}\lambda_{P_{ij}}=\left(N-2\right)\left(\lambda_{P_{1}}+\lambda_{P_{2}}\right)+\sum_{i=3}^{N}\left(N+1-2i\right)\lambda_{P_{i}}$.
And then the wave function for $z_{2}=z_{1}+\varepsilon$ can be rewritten
as
\begin{align}
\psi\left(z_{1},\cdots,z_{N}\right) & \approx\sum_{P}\left(-1\right)^{P}\lambda_{P_{12}}\left(\frac{\left|l_{p}\right|}{\textrm{i}}+l_{p}^{2}\left[\frac{\xi_{p}}{\textrm{i}}\lambda_{P_{12}}^{2}-\left(N-2\right)\left(\lambda_{P_{1}}+\lambda_{P_{2}}\right)-\sum_{i=3}^{N}\left(N+1-2i\right)\lambda_{P_{i}}\right]\right)e^{\mathrm{i}\sum_{j=1}^{N}\lambda_{P_{j}}z_{j}}\nn\\
 & =\psi^{\left(1\right)}\left(z_{1},\cdots,z_{N}\right)\left|l_{p}\right|+\psi^{\left(2\right)}\left(z_{1},\cdots,z_{N}\right)l_{p}^{2},
\end{align}
in which the wave function to the first order and the second order
of $\left|l_{p}\right|$ are given by
\begin{align}
\psi^{\left(1\right)}\left(z_{1},\cdots,z_{N}\right) & =\sum_{P}\left(-1\right)^{P}\frac{\lambda_{P_{12}}}{\textrm{i}}e^{\mathrm{i}\sum_{j=1}^{N}\lambda_{P_{j}}z_{j}}\nn\\
 & =\sum_{P}\left(-1\right)^{P}\left(\partial_{z_{2}}-\partial_{z_{1}}\right)e^{\mathrm{i}\sum_{j=1}^{N}\lambda_{P_{j}}z_{j}}\nn\\
 & =\left(\partial_{z_{2}}-\partial_{z_{1}}\right)\psi^{\left(0\right)}\left(z_{1},\cdots,z_{N}\right),
\end{align}
\begin{align}
\psi^{\left(2\right)}\left(z_{1},\cdots,z_{N}\right) & =\sum_{P}\left(-1\right)^{P}\lambda_{P_{12}}\left[\frac{\xi_{p}}{\textrm{i}}\lambda_{P_{12}}^{2}-\left(N-2\right)\left(\lambda_{P_{1}}+\lambda_{P_{2}}\right)-\sum_{i=3}^{N}\left(N+1-2i\right)\lambda_{P_{i}}\right]e^{\mathrm{i}\sum_{j=1}^{N}\lambda_{P_{j}}z_{j}}\nn\\
 & =\sum_{P}\textrm{i}\left(-1\right)^{P}\left(\partial_{z_{2}}-\partial_{z_{1}}\right)\left[\textrm{i}\xi_{p}\left(\partial_{z_{2}}-\partial_{z_{1}}\right)^{2}+\textrm{i}\left(N-2\right)\left(\partial_{z_{2}}+\partial_{z_{1}}\right)+\textrm{i}\sum_{i=3}^{N}\left(N+1-2i\right)\partial_{z_{i}}\right]e^{\mathrm{i}\sum_{j=1}^{N}\lambda_{P_{j}}z_{j}}\nn\\
 & =-\left[\xi_{p}\left(\partial_{z_{2}}-\partial_{z_{1}}\right)^{2}+\left(N-2\right)\left(\partial_{z_{2}}+\partial_{z_{1}}\right)+\sum_{i=3}^{N}\left(N+1-2i\right)\partial_{z_{i}}\right]\psi^{\left(1\right)}\left(z_{1},\cdots,z_{N}\right),
\end{align}
with the zeroth order wave function $\psi^{\left(0\right)}\left(z_{1},\cdots,z_{N}\right)=\sum_{P}\left(-1\right)^{P}e^{\mathrm{i}\sum_{j=1}^{N}\lambda_{P_{j}}z_{j}}$. 

The correlation function without normalization for infinitesimal $y_{1},y_{2};z_{1},z_{2}$
is calculated by above wave function
\begin{align}
 & g_{2}\left(y_{1},y_{2};z_{1},z_{2}\right)\nn\\
 & =\int_{0}^{L}dz_{3}\cdots\int_{0}^{L}dz_{N}\psi^{*}\left(y_{1},y_{2},z_{3},\cdots\right)\psi\left(z_{1},\cdots,z_{N}\right)\nn\\
 & =\left(N-2\right)!\int_{0\leq z_{3}\leq\cdots\leq z_{N}\leq L}\psi^{*}\left(y_{1},y_{2},z_{3},\cdots\right)\psi\left(z_{1},\cdots,z_{N}\right)\nn\\
 & \approx\left(N-2\right)!\rho^{\left(\Delta\right)}l_{p}^{2}-\left(N-2\right)!\left|l_{p}^{3}\right|\left[\xi_{p}\left[\left(\partial_{z_{2}}-\partial_{z_{1}}\right)^{2}+\left(\partial_{y_{2}}-\partial_{y_{1}}\right)^{2}\right]+\left(N-2\right)\left(\partial_{z_{2}}+\partial_{z_{1}}+\partial_{y_{2}}+\partial_{y_{1}}\right)\right]\rho^{\left(\Delta\right)}\nn\\
 & -\left(N-2\right)!\left|l_{p}^{3}\right|\sum_{i=3}^{N}\left(N+1-2i\right)\int_{0}^{L}\partial_{z_{i}}\rho_{i}^{\left(\Delta\right)}\left(z_{i}\right)dz_{i}.
\end{align}
in which
\begin{equation}
\rho^{\left(\Delta\right)}\equiv\int_{0\leq z_{3}\leq\cdots\leq z_{N}\leq L}\psi^{\left(1\right)*}\left(y_{1},y_{2},z_{3},\cdots\right)\psi^{\left(1\right)}\left(z_{1},\cdots,z_{N}\right)dz_{3}\cdots dz_{N},
\end{equation}
\begin{equation}
\rho_{l}^{\left(\Delta\right)}\left(z_{l}\right)\equiv\int_{0\leq z_{3}\leq\cdots\leq z_{N}\leq L}\psi^{\left(1\right)*}\left(y_{1},y_{2},z_{3},\cdots\right)\psi^{\left(1\right)}\left(z_{1},\cdots,z_{N}\right)dz_{3}\cdots dz_{l-1}dz_{l+1}\cdots dz_{N},
\end{equation}
with $3\le l\le N$. For the first term in the correlation function,
we define the following derivatives
\begin{align}
g_{2}^{\left(2\right)}\left(y_{1},y_{2};z_{1},z_{2}\right) & \equiv\left(N-2\right)!\rho^{\left(\Delta\right)}l_{p}^{2}\nn\\
 & =l_{p}^{2}\left(\partial_{y_{2}}-\partial_{y_{1}}\right)\left(\partial_{z_{2}}-\partial_{z_{1}}\right)\left(N-2\right)!\int_{0\leq z_{3}\leq\cdots\leq z_{N}\leq L}\psi^{\left(0\right)*}\left(y_{1},y_{2},z_{3},\cdots\right)\psi^{\left(0\right)}\left(z_{1},\cdots,z_{N}\right)dz_{3}\cdots dz_{N}\nn\\
 & =l_{p}^{2}\left(\partial_{y_{2}}-\partial_{y_{1}}\right)\left(\partial_{z_{2}}-\partial_{z_{1}}\right)g_{2}^{\left(0\right)}\left(y_{1},y_{2};z_{1},z_{2}\right)
\end{align}
with $g_{2}^{\left(0\right)}\left(y_{1},y_{2};z_{1},z_{2}\right)=(N-2)\int_{0}^{L}dz_{3}\cdots\int_{0}^{L}dz_{N}\left[\psi^{\left(0\right)}\left(y_{1},y_{2},z_{3},\cdots\right)\right]^{*}\psi^{\left(0\right)}\left(z_{1},\cdots,z_{N}\right)$.
The last term in the correlation function can be proved to be zero
with
\begin{equation}
\int_{0}^{L}\partial_{z_{i}}\rho_{i}^{\left(\Delta\right)}\left(z_{i}\right)dz_{i}=\rho_{i}^{\left(\Delta\right)}\left(L\right)-\rho_{i}^{\left(\Delta\right)}\left(0\right)=0.
\end{equation}
At last, one can obtain the correlation function
\begin{equation}
g_{2}\left(y_{1},y_{2};z_{1},z_{2}\right)=\left\{ 1-\left|l_{p}\right|\xi_{p}\left[\left(\partial_{z_{2}}-\partial_{z_{1}}\right)^{2}+\left(\partial_{y_{2}}-\partial_{y_{1}}\right)^{2}\right]\right\} g_{2}^{\left(2\right)}\left(y_{1},y_{2};z_{1},z_{2}\right).
\end{equation}

The Bethe ansatz equations (\ref{eq:BA-equations2})   for weak atrractive interaction ($\left|l_{p}\right|\ll L$)
exhibit  the asymptotic solution $\lambda_{i}=\lambda_{i}^{F}\alpha$ for total momentum
$\sum_{i=1}^{N}\lambda_{i}=0$, where $\lambda_{i}^{F}=2\pi J_{i}/L$,
$\alpha=1-2\left|l_{p}\right|n$, and the $J_{i}$ are (half-)integers
satisfying $J_{1}<J_{2}<...<J_{N}$. Under the scaling $z_{i}=z_{i}^{F}/\alpha$,
the zeroth order correlation function is simplified to
\begin{equation}
g_{2}^{0}\left(y_{1},y_{2};z_{1},z_{2}\right)\approx\alpha^{2-N}\int_{0}^{L}\psi^{F}\left(y_{1}^{F},y_{2}^{F},...,z_{N}^{F}\right)\psi^{F}\left(z_{1}^{F},z_{2}^{F},...,z_{N}^{F}\right)dz_{3}^{F}...dz_{N}^{F}
\end{equation}
with wave function of ideal Fermi gas $\psi^{F}\left(z_{1}^{F},z_{2}^{F},...,z_{N}^{F}\right)=\sum_{Q}\left(-1\right)^{Q}e^{\mathrm{i}\sum_{j=1}^{N}k_{Q_{j}}^{F}z_{j}^{F}}$.
Meanwhile using the normalization factor\cite{correlation_Nandani2016}
\begin{eqnarray}
\int_{0}^{L}dz_{1}...\int_{0}^{L}dz_{N}\left|\Psi\left(z_{1},z_{2},...,z_{N}\right)\right|^{2} & = & \alpha^{1-N}\int_{0}^{L}dz_{1}...\int_{0}^{L}dz_{N}\left|\psi^{\left(0\right)}\left(z_{1},z_{2},...,z_{N}\right)\right|^{2}\nn\\
 & \approx & \alpha^{1-N}\int_{0}^{L}dz_{1}^{F}...\int_{0}^{L}dz_{N}^{F}\left|\psi^{F}\left(z_{1}^{F},...,z_{N}^{F}\right)\right|^{2},
\end{eqnarray}
the zeroth order correlation function with normalization factor has
the following form
\begin{equation}
g_{2}^{0}\left(y_{1},y_{2};z_{1},z_{2}\right)=\alpha N\left(N-1\right)\frac{\int_{0}^{L}\psi^{F}\left(y_{1}^{F},y_{2}^{F},...,z_{N}^{F}\right)\psi^{F}\left(z_{1}^{F},z_{2}^{F},...,z_{N}^{F}\right)dz_{3}^{F}...dz_{N}^{F}}{\int_{0}^{L}dz_{1}^{F}...\int_{0}^{L}dz_{N}^{F}\left|\psi^{F}\left(z_{1}^{F},...,z_{N}^{F}\right)\right|^{2}}.
\end{equation}
Since $\psi^{F}\left(z_{1}^{F},z_{2}^{F},...,z_{N}^{F}\right)$ is
a Slater determinant,  thus one can obtain the following result  by the  Wick's
theorem
\begin{equation}
g_{2}^{0}\left(y_{1},y_{2};z_{1},z_{2}\right)=\alpha\left[G\left(y_{1}^{F},z_{1}^{F}\right)G\left(y_{2}^{F},z_{2}^{F}\right)-G\left(y_{1}^{F},z_{2}^{F}\right)G\left(y_{2}^{F},z_{1}^{F}\right)\right],
\end{equation}
where the single particle reduced density matrix of ideal fermions
is given by $G\left(y,z\right)=\frac{1}{L}\sum_{i=1}^{N}\exp\left[-\textrm{i}\lambda_{i}^{F}\left(y-z\right)\right]$.
By using the above result, the two-body correlation function
has the following form
\begin{align}
g_{2}^{\left(2\right)}\left(y_{1},y_{2};z_{1},z_{2}\right) & =l_{p}^{2}\left(\partial_{y_{2}}-\partial_{y_{1}}\right)\left(\partial_{z_{2}}-\partial_{z_{1}}\right)g_{2}^{\left(0\right)}\left(y_{1},y_{2};z_{1},z_{2}\right)\nn\\
 & =l_{p}^{2}\alpha^{3}\left(\partial_{y_{2}^{F}}-\partial_{y_{1}^{F}}\right)\left(\partial_{z_{2}^{F}}-\partial_{z_{1}^{F}}\right)\left[G\left(y_{1}^{F},z_{1}^{F}\right)G\left(y_{2}^{F},z_{2}^{F}\right)-G\left(y_{1}^{F},z_{2}^{F}\right)G\left(y_{2}^{F},z_{1}^{F}\right)\right]\nn\\
 & =l_{p}^{2}\frac{1}{L^{2}}\sum_{i,j=1}^{N}\alpha^{3}\left(\lambda_{i}^{F}-\lambda_{j}^{F}\right)^{2}X\left(\lambda_{i}^{F},\lambda_{j}^{F}\right),
\end{align}
where $X\left(\lambda_{i}^{F},\lambda_{j}^{F}\right)=\exp\left[-\textrm{i}\lambda_{i}^{F}\left(y_{1}^{F}-z_{1}^{F}\right)\right]\exp\left[-\textrm{i}\lambda_{j}^{F}\left(y_{2}^{F}-z_{2}^{F}\right)\right]+\exp\left[-\textrm{i}\lambda_{i}^{F}\left(y_{1}^{F}-z_{2}^{F}\right)\right]\exp\left[-\textrm{i}\lambda_{j}^{F}\left(y_{2}^{F}-z_{1}^{F}\right)\right]$.
Finally, we have the two-body correlation function (\ref{eq:g2-Green}) in the main text, namely
\begin{equation}
g_{2}\left(y_{1},y_{2};z_{1},z_{2}\right)=l_{p}^{2}\frac{1}{L^{2}}\sum_{i,j=1}^{N}\left[\alpha^{3}\left(\lambda_{i}^{F}-\lambda_{j}^{F}\right)^{2}+2\left|l_{p}\right|\alpha^{5}\xi_{p}\left(\lambda_{i}^{F}-\lambda_{j}^{F}\right)^{4}\right]X\left(\lambda_{i}^{F},\lambda_{j}^{F}\right).
\end{equation}

 \section{P-wave contacts: $C_2$ and $C_4$ }
 
 For the low temperature, $A/T\gg1$, according to Sommerfeld expansion,  the function $f_{i}=\int_{-\infty}^{\infty}dx\mathcal{N}\left(x\right)x^{i}=\frac{1}{\left(2\pi n\alpha\right)^{i+1}}\int_{0}^{\infty}dy\frac{y^{\left(i-1\right)/2}}{1+e^{\left(y-A\right)/T}}$
is expanded with respect to the temperature
\begin{equation}
f_{i}=\frac{1}{\left(2\pi n\alpha\right)^{i+1}}\left[\frac{2}{i+1}A^{\left(i+1\right)/2}+\frac{\pi^{2}\left(i-1\right)}{12}T^{2}A^{\left(i-3\right)/2}+...\right].
\end{equation}
From the normalization condition $f_{0}=1$, the effective chemical
potential $A$ can be solved by iteration method
\begin{equation}
A\approx\left(\pi n\alpha\right)^{2}\left(1+\frac{1}{12\pi^{2}}\frac{\tau^{2}}{\alpha^{4}}\right)
\end{equation}
with $\tau=T/n^{2}$. And put the effective chemical potential into
$f_{i}$, $f_{i}$ is expressed with $\tau$ and $\alpha$
\begin{equation}
f_{i}=\frac{1}{\left(i+1\right)2^{i}}\left[1+\frac{i\left(i+1\right)}{24\pi^{2}}\frac{\tau^{2}}{\alpha^{4}}+...\right]
\end{equation}
and then one obtain the contact $C_{2}$ and $C_{4}$
\begin{equation}
C_{2}=\frac{8}{3}\left|l_{p}\right|^{2}\pi n^{3}N\left[\left(1-6\left|l_{p}\right|n\right)+\left(1+2\left|l_{p}\right|n\right)\frac{1}{4\pi^{2}}\tau^{2}+\frac{16}{5}\left|l_{p}\right|\xi_{p}\left(\pi n\right)^{2}\left(1+\frac{5}{8\pi^{2}}\tau^{2}\right)\right],
\end{equation}
\begin{equation}
C_{4}=\frac{8}{3}\left|l_{p}\right|^{2}\pi^{3}n^{5}N\left[\left(1-10\left|l_{p}\right|n\right)+\left(1-2\left|l_{p}\right|n\right)\frac{3}{4\pi^{2}}\tau^{2}+\frac{132}{35}\left|l_{p}\right|\xi_{p}\left(\pi n\right)^{2}\left(1+\frac{14}{11\pi^{2}}\tau^{2}\right)\right].
\end{equation}

For the relative high temperature which requires effective fugacity
$Z_{A}\equiv\exp\left(A/T\right)\ll1$, the integral $f_{i}$ is expanded
by effective fugacity
\begin{align}
f_{i} & \approx\frac{1}{\left(2\pi n\alpha\right)^{i+1}}\int_{0}^{\infty}dyy^{\left(i-1\right)/2}Z_{A}e^{-y/T}\left(1-Z_{A}e^{-y/T}\right)\nn\\
 & =\frac{T^{\left(i+1\right)/2}}{\left(2\pi n\alpha\right)^{i+1}}\Gamma\left(\frac{i+1}{2}\right)Z_{A}\left(1-2^{-\left(i+1\right)/2}Z_{A}\right).
\end{align}
From the normalization condition $f_{0}=1$, the effective fugacity
can be solved by iteration method
\begin{equation}
Z_{A}\approx\frac{2\sqrt{\pi}\alpha}{\tau^{1/2}}\left(1+\frac{\sqrt{2\pi}\alpha}{\tau^{1/2}}\right),
\end{equation}
and then one obtain the contact $C_{2}$ and $C_{4}$
\begin{equation}
C_{2}=\frac{4}{\pi}l_{p}^{2}Nn^{3}\tau\left[\left(1-2\left|l_{p}\right|n\right)+\left(1-4\left|l_{p}\right|n\right)\frac{\sqrt{2\pi}}{2}\tau^{-1/2}+6\left|l_{p}\right|\xi_{p}T\left(1+\frac{7\sqrt{2\pi}}{8}\tau^{-1/2}\right)\right],
\end{equation}
\begin{equation}
C_{4}=\frac{9}{\pi}l_{p}^{2}Nn^{3}\tau^{2}\left[\left(1-2\left|l_{p}\right|n\right)+\left(1-4\left|l_{p}\right|n\right)\frac{19\sqrt{2\pi}}{24}\tau^{-1/2}+\frac{26}{3}\left|l_{p}\right|\xi_{p}T\left(1+\frac{445\sqrt{2\pi}}{416}\tau^{-1/2}\right)\right].
\end{equation}

 \section{The thermodynamic Bethe Ansatz equations}
 The thermodynamics Bethe Ansatz equation for p-wave fermions in one
dimension can be written as

\begin{equation}
\epsilon(\lambda)=\frac{\hbar^{2}\lambda^{2}}{2m}-\mu-\frac{T}{2\pi}\int_{-\infty}^{\infty}K\left(k-\lambda\right)\mathrm{ln}\left\{ 1+\mathrm{exp}\left[-\epsilon\left(k\right)/T\right]\right\} dk\label{eq:TBA}
\end{equation}
with $K(x)=\frac{2|l_{p}|(|l_{p}|\xi_{p}x^{2}+1)}{(|l_{p}|\xi_{p}x^{2}-1)^{2}+l_{p}^{2}x^{2}}$.
The pressure is defined as 
\begin{equation}
P=\frac{T}{2\pi}\int_{-\infty}^{\infty}\mathrm{ln}\left\{ 1+\mathrm{exp}\left[-\epsilon\left(\lambda\right)/T\right]\right\} d\lambda.\label{Pressure}
\end{equation}
The grand canonical potential, the particle density per length and
the contact are given by $\Omega=-PL$, $n=\frac{\partial P}{\partial\mu}$
and $C_{2}=\frac{2L}{\pi}l_{p}^{2}\frac{\partial P}{\partial|l_{p}|}$.

For weak interaction, $\left|l_{p}\right|<1/n$, the kernel in the
integral function of equation (\ref{eq:TBA})is expanded up to the
second order of $\left|l_{p}\right|$
\begin{equation}
K(k-\lambda)\approx2\left|l_{p}\right|+6\xi_{p}\left|l_{p}\right|^{2}\lambda^{2}+12\xi_{p}\left|l_{p}\right|^{2}\lambda k+6\xi_{p}\left|l_{p}\right|^{2}k^{2},
\end{equation}
and then TBA equation is simplified as
\begin{equation}
\epsilon\left(\lambda\right)=B\lambda^{2}-A
\end{equation}
with the effective chemical potential $A=\mu+2\left|l_{p}\right|Q_{0}+6Q_{2}\xi_{p}\left|l_{p}\right|^{2}$
with the notations $B=1-6Q_{0}\xi_{p}\left|l_{p}\right|^{2}$ and
$Q_{j}=\frac{T}{2\pi}\int_{-\infty}^{\infty}k^{j}\mathrm{ln}\left\{ 1+\mathrm{exp}\left[-\frac{\epsilon\left(k\right)}{T}\right]\right\} dk$.
Using integral by part, $Q_{j}$ can be expressed by polylogarithm
\begin{equation}
Q_{j}=-\left(\frac{T}{B}\right)^{\frac{j+1}{2}}\frac{T}{\pi\left(j+1\right)}\Gamma\left(\frac{j+3}{2}\right)\mathrm{Li}_{\frac{j+3}{2}}\left(-\mathrm{e}^{A/T}\right),
\end{equation}
and using the properties of polylogarithm $\textrm{Li}_{i}\left(-\mathrm{e}^{A/T}\right)$,
one obtain Pressure, particle density and contact with the following
forms
\begin{equation}
P=-\frac{T^{3/2}}{2\sqrt{\pi}}\mathrm{Li}_{\frac{3}{2}}\left(-\mathrm{e}^{A/T}\right)\left(1-\frac{3T^{3/2}}{2\sqrt{\pi}}\mathrm{Li}_{\frac{3}{2}}\left(-\mathrm{e}^{A/T}\right)\xi_{p}\left|l_{p}\right|^{2}\right),
\end{equation}
\begin{equation}
n=-\frac{T^{1/2}}{2\sqrt{\pi}}\mathrm{Li}_{\frac{1}{2}}\left(-\mathrm{e}^{A/T}\right)\left(1+2\left|l_{p}\right|n\right),
\end{equation}
\begin{equation}
C_{2}=\frac{2L}{\pi}l_{p}^{2}n\left[2P-\left(\frac{3T^{5/2}}{\sqrt{\pi}}\mathrm{Li}_{\frac{5}{2}}\left(-\mathrm{e}^{A/T}\right)-\frac{6P^{2}}{n}\right)\xi_{p}\left|l_{p}\right|\right],
\end{equation}
in which the expression of particle density is equivalent to previous
normalization condition of modified Fermi-Dirac distribution function,
and contact $C_{2}$ is also consistent with previous result from
many-body wave function.

In the low temperature limit($T\ll T_{d}$), according to Sommerfeld
expansion, $Q_{j}$ is expanded with respect to temperature
\begin{equation}
Q_{j}\approx\left(\frac{1}{B}\right)^{\frac{j+1}{2}}\frac{1}{\pi\left(j+1\right)}\left[\frac{2}{j+3}A^{\left(j+3\right)/2}+\frac{j+1}{2}\frac{\pi^{2}}{6}T^{2}A^{\left(j-1\right)/2}\right].
\end{equation}
When the iteration method is applied, the pressure is rewritten as
\begin{equation}
P=\frac{2}{3\pi}\mu^{3/2}\left\{ 1+2\frac{\sqrt{\mu}}{\pi}\left|l_{p}\right|+\left(\frac{14}{3}+\frac{16\pi}{5}\sqrt{\mu}\xi_{p}\right)\frac{\mu}{\pi^{2}}l_{p}^{2}+\frac{\pi^{2}}{8}\frac{T^{2}}{\mu^{2}}\left[1+\frac{4}{3}\frac{\sqrt{\mu}}{\pi}\left|l_{p}\right|+\left(\frac{10}{3}+\frac{48\pi}{5}\sqrt{\mu}\xi_{p}\right)\frac{\mu}{\pi^{2}}l_{p}^{2}\right]\right\} .\label{eq:P_lowT}
\end{equation}
According to formula $C_{2}=\frac{2L}{\pi}l_{p}^{2}\frac{\partial P}{\partial|l_{p}|}$,
the contact is
\begin{equation}
C_{2}=\frac{8Ll_{p}^{2}}{3\pi^{3}}\mu^{2}\left\{ 1+\left(\frac{14}{3}+\frac{16\pi}{5}\sqrt{\mu}\xi_{p}\right)\frac{\sqrt{\mu}}{\pi}\left|l_{p}\right|+\frac{\pi^{2}}{12}\frac{T^{2}}{\mu^{2}}\left[1+\left(5+\frac{72\pi}{5}\sqrt{\mu}\xi_{p}\right)\frac{\sqrt{\mu}}{\pi}\left|l_{p}\right|\right]\right\} ,
\end{equation}
in which chemical potential can be expressed by particle density with
the form
\begin{equation}
\mu=\pi^{2}n^{2}\left[1-\frac{16}{3}n\left|l_{p}\right|+\frac{1}{12}\frac{T^{2}}{\pi^{2}n^{4}}\right].
\end{equation}
At last one obtain the contact
\begin{equation}
C_{2}=\frac{8}{3}\pi l_{p}^{2}n^{3}N\left[\left(1-6|l_{p}|n\right)+\frac{1}{4}\frac{T^{2}}{\pi^{2}n^{4}}\left(1+2|l_{p}|n\right)+\frac{16}{5}n^{2}\pi^{2}\xi_{p}\left|l_{p}\right|\left(1+\frac{5}{8}\frac{T^{2}}{\pi^{2}n^{4}}\right)\right]
\end{equation}
This result is consistent with that from many-body wave function.

For the relative high temperature, the fugacity $Z=\mathrm{e}^{\mu/T}\ll1$,
$Q_{j}$ can be expanded by fugacity
\begin{equation}
Q_{j}=\left(\frac{T}{B}\right)^{\frac{j+1}{2}}\frac{T}{\pi\left(j+1\right)}\Gamma\left(\frac{j+3}{2}\right)\mathrm{e}^{A/T}\left[1-2^{-\left(j+3\right)/2}\mathrm{e}^{A/T}+\mathrm{e}^{2A/T}3^{-\left(j+3\right)/2}-\mathrm{e}^{3A/T}4^{-\left(j+3\right)/2}\right].
\end{equation}
When the iteration method is applied, the pressure is rewritten as
\begin{equation}
P=\frac{1}{2}\frac{ZT^{3/2}}{\sqrt{\pi}}\left[1-\frac{\sqrt{2}}{4}Z+\frac{Z\sqrt{T}}{\sqrt{\pi}}\left|l_{p}\right|\left(1-\frac{3\sqrt{2}}{4}Z\right)+\frac{3}{2}\frac{Z^{2}T}{\text{\ensuremath{\pi}}}l_{p}^{2}\left(1-\frac{4\sqrt{2}}{3}Z\right)+\frac{3ZT^{3/2}}{\sqrt{\text{\ensuremath{\pi}}}}l_{p}^{2}\xi_{p}\left(1-\frac{9\sqrt{2}}{16}Z\right)\right].
\end{equation}
According to formula $C_{2}=\frac{2L}{\pi}l_{p}^{2}\frac{\partial P}{\partial|l_{p}|}$,
the contact is
\begin{equation}
C_{2}=\frac{Ll_{p}^{2}Z^{2}T^{2}}{\pi^{2}}\left\{ 1-\frac{3}{4}\sqrt{2}Z+\left(1-\frac{4\sqrt{2}}{3}Z\right)\frac{3Z\sqrt{T}}{\sqrt{\pi}}|l_{p}|+6|l_{p}|\xi_{p}T\left(1-\frac{9}{16}\sqrt{2}Z\right)\right\} ,
\end{equation}
in which fugacity can be expressed in particle density
\begin{equation}
Z=\frac{2\sqrt{\pi}n}{\sqrt{T}}\left(1+\frac{\sqrt{2\pi}n}{\sqrt{T}}\right)-\left(1+\frac{7\sqrt{2\pi}n}{4\sqrt{T}}\right)\frac{8\sqrt{\pi}n}{\sqrt{T}}|l_{p}|n.
\end{equation}
At last, one obtain contact
\begin{equation}
C_{2}=\frac{4l_{p}^{2}TnN}{\pi}\left[1-2n|l_{p}|+\left(1-4n|l_{p}|\right)\frac{\sqrt{2\pi}}{2}\frac{n}{\sqrt{T}}+6|l_{p}|\xi_{p}T\left(1+\frac{7\sqrt{2\pi}}{8}\frac{n}{\sqrt{T}}\right)\right],
\end{equation}
which is also consistent with that from many-body wave function.
 
 \section{Strong coupling limit} 
 The energy of system composed of $N$ spinless fermions can be expressed
as $E=\sum_{i}^{N}\lambda_{i}^{2}$, and quasi-momentum $\lambda_{i}$
satisfy BA equation for real $\lambda_{i}$,
\begin{equation}
\lambda_{i}L=2\pi n_{i}-\sum_{j=1}^{N}\theta\left(\lambda_{i}-\lambda_{j}\right).\label{eq:BA1}
\end{equation}
Here phase shift $\theta\left(\lambda\right)$ is a monotonic antisymmetric
function defined by
\begin{equation}
\theta\left(\lambda\right)=2\arg\left(\textrm{i}\lambda-\xi_{p}\lambda^{2}+1/\left|l_{p}\right|\right)\label{eq:theta1}
\end{equation}
and $n_{i}=i-\left(N+1\right)/2$ for ground state. In the strong
attractive limit $\left|l_{p}\right|\gg1/L$, and we assume that $\left|-\xi_{p}\lambda^{2}+1/\left|l_{p}\right|\right|\ll\left|\lambda\right|$,
and then phase shift is approach to $\pm\pi$ with $\left(\lambda\right)\approx\pi\lambda/\left|\lambda\right|-2\left(\frac{x}{\lambda}-\xi_{p}\lambda\right)$
and $x\equiv1/\left|l_{p}\right|$. And then BA equation becomes new
form
\begin{equation}
\lambda_{i}L=\sum_{j=1\left(j\neq i\right)}^{N}\left[\frac{2x}{\lambda_{i}-\lambda_{j}}-2\xi_{p}\left(\lambda_{i}-\lambda_{j}\right)\right]\label{eq:BA4}
\end{equation}
For the ground state, the total momentum is zero, which means $\sum_{j}\lambda_{j}=0$,
so BA equation can be simplified as the new form
\begin{equation}
\left(L+2N\xi_{p}\right)\lambda_{i}=\sum_{j=1\left(j\neq i\right)}^{N}\frac{2x}{\lambda_{i}-\lambda_{j}}\label{eq:BA6}
\end{equation}
The solution of above equation has the following form$\lambda_{i}=\sqrt{\frac{2x}{L+2N\xi_{p}}}\beta_{i}$,
with constant $\beta_{i}$ which is determined by a set of equations
$\beta_{i}=\sum_{j=1\left(j\neq i\right)}^{N}\frac{1}{\beta_{i}-\beta_{j}}$.
The previous assumation $\left|-\xi_{p}\lambda^{2}+1/\left|l_{p}\right|\right|\ll\left|\lambda\right|$
requires that $\left|l_{p}\right|\gg L+2N\xi_{p}$, which is strong
coupling condition for above solution. With above solution, the energy
of system is
\begin{equation}
E=\frac{2/\left|l_{p}\right|}{L+2N\xi_{p}}\sum_{i}\beta_{i}^{2}=\frac{N\left(N-1\right)/\left|l_{p}\right|}{L+2N\xi_{p}}.\label{eq:E6}
\end{equation}

The coefficients in the wavefunction can be written as
\begin{eqnarray}
a\left(P\right) & = & \left(-1\right)^{P}\prod_{i<j}\left(\exp\left(-\textrm{i}\theta\left(\lambda_{P_{i}}-\lambda_{P_{j}}\right)\right)\right)^{1/2}\nn\\
 & = & \exp\left(\textrm{i}\frac{\pi}{2}\left(1-\left(-1\right)^{P}\right)-\frac{\textrm{i}}{2}\sum_{i<j}\theta\left(\lambda_{P_{i}}-\lambda_{P_{j}}\right)\right).
\end{eqnarray}
In the strong attractive interaction limit, $\lambda_{i}=\frac{H\left(\xi_{p}\right)}{\sqrt{\left|l_{p}\right|}}\beta_{i}$
with $H\left(\xi_{p}\right)=\sqrt{\frac{2}{L+2N\xi_{p}}}$ , and phase
shift $\theta\left(\lambda\right)\approx\frac{\lambda}{\left|\lambda\right|}\pi-2F\left(\xi_{p},\beta\right)\frac{1}{\sqrt{\left|l_{p}\right|}}$,
with $F\left(\xi_{p},\beta\right)=\frac{1}{\sqrt{\frac{2}{L+2N\xi_{p}}}\beta}\left(1-\frac{2\xi_{p}}{L+2N\xi_{p}}\beta^{2}\right)$,
and $F\left(\xi_{p},-\beta\right)=-F\left(\xi_{p},\beta\right)$.
The coefficients has the following form
\begin{equation}
a\left(P\right)=\exp\left(\frac{\textrm{i}}{\sqrt{\left|l_{p}\right|}}B\left(P\right)\right)
\end{equation}
with $B\left(P\right)=\sum_{i<j}F\left(\xi_{p},\beta_{P_{i}}-\beta_{P_{j}}\right)$.
The wavefunction in the domain $z_{1}\leq z_{2}\leq...\leq z_{N}$
can be simplified as
\begin{eqnarray}
\psi\left(z_{1},z_{2},...,z_{N}\right) & = & \sum_{P}\exp\left(\frac{\textrm{i}}{\sqrt{\left|l_{p}\right|}}B\left(P\right)\right)\exp\left(\frac{\mathrm{i}}{\sqrt{\left|l_{p}\right|}}H\left(\xi_{p}\right)\sum_{j=1}^{N}\beta_{P_{j}}z_{j}\right)\nn\\
 & \approx & N!-\frac{Y\left(z_{1},z_{2},...,z_{N},\xi_{p}\right)}{\left|l_{p}\right|}+\cdots
\end{eqnarray}
with $Y=\frac{1}{2}\sum_{P}\left[B\left(P\right)+H\left(\xi_{p}\right)\sum_{j=1}^{N}\beta_{P_{j}}z_{j}\right]^{2}$.
The correlation function can be written as
\begin{align}
& g_{2}(y_{1},y_{2};z_{1},z_{2})\nn\\
& =\frac{N\left(N-1\right)\int_{0}^{L}dz_{3}...\int_{0}^{L}dz_{N}\psi^{*}\left(y_{1},y_{2},z_{3},...\right)\psi\left(z_{1},z_{2},z_{3},...\right)}{\int_{0}^{L}dz_{1}...\int_{0}^{L}dz_{N}\left|\Psi\left(z_{1},z_{2},...,z_{N}\right)\right|^{2}}\nn\\
 & \approx\frac{N\left(N-1\right)\left[\left(N-2\right)!\right]N!}{\left(N!\right)^{2}L^{N}}\int_{0\leq z_{3}\leq...\leq z_{N}\leq L}dz_{3}...dz_{N}\left[N!-\frac{Y\left(y_{1},y_{2},z_{3},...,z_{N},\xi_{p}\right)}{\left|l_{p}\right|}-\frac{Y\left(z_{1},z_{2},...,z_{N},\xi_{p}\right)}{\left|l_{p}\right|}\right],
\end{align}
in which we suppose that $y_{1},y_{2},z_{1},z_{2}$ is close to zero.
The leading term of contact $C_{2}$, $C_{4}^{r}$, $C_{4}^{c}$ and
$C_{4}$ have the following results
\begin{align}
C_{2} & =\frac{2L}{\pi}g_{2}\left(0,\varepsilon;0,\varepsilon\right)\nn\\
 & \approx\frac{2L}{\pi}\frac{N\left(N-1\right)\left(N!\right)^{2}L^{N-2}}{\left(N!\right)^{2}L^{N}}\nn\\
 & =\frac{2}{\pi}\frac{N\left(N-1\right)}{L}.
\end{align}
\begin{align}
C_{4}^{r} & =-\frac{L}{\pi}\left[\textrm{Re}\left(\partial_{z_{2}}-\partial_{z_{1}}\right)^{2}g_{2}\left(y_{2}-\varepsilon,y_{2};z_{1},z_{2}\right)\right]_{y_{2}=z_{2}=z_{1}+\varepsilon}\nn\\
 & \approx-\frac{L}{\pi}\frac{\frac{-1}{\left|l_{p}\right|}\left(N!\right)^{2}L^{N-2}\frac{2N}{L+2N\xi_{p}}N\left(N-1\right)}{\left(N!\right)^{2}L^{N}}\nn\\
 & =\frac{1}{\pi}\frac{1}{\left|l_{p}\right|}\frac{2nN\left(N-1\right)}{L+2N\xi_{p}},
\end{align}
\begin{align}
C_{4}^{c} & =\frac{L}{2\pi}\left[\partial_{y_{2}}\partial_{z_{2}}g_{2}\left(y_{2}-\varepsilon,y_{2};z_{2}-\varepsilon,z_{2}\right)-2\textrm{Re}\partial_{z_{2}}^{2}g_{2}\left(y_{2}-\varepsilon,y_{2};z_{2}-\varepsilon,z_{2}\right)\right]_{y_{2}=z_{2}}\nn\\
 & \approx0+N^{2}\left(N-1\right)\frac{L}{2\pi N}\left(-2\right)\frac{-1}{\left|l_{p}\right|}\left(N!\right)^{2}L^{N-2}\frac{2}{L+2N\xi_{p}}\left(N-2\right)/\left[\left(N!\right)^{2}L^{N}\right]\nn\\
 & =\frac{1}{\left|l_{p}\right|}\frac{2N\left(N-1\right)\left(N-2\right)}{\pi L\left(L+2N\xi_{p}\right)},
\end{align}
\begin{equation}
C_{4}=C_{4}^{c}+C_{4}^{r}=\frac{1}{\left|l_{p}\right|}\frac{4N\left(N-1\right)^{2}}{\pi L\left(L+2N\xi_{p}\right)}.
\end{equation}

\bibliography{ref.bib}

\begin{thebibliography}{54}
\expandafter\ifx\csname natexlab\endcsname\relax\def\natexlab#1{#1}\fi
\expandafter\ifx\csname bibnamefont\endcsname\relax
  \def\bibnamefont#1{#1}\fi
\expandafter\ifx\csname bibfnamefont\endcsname\relax
  \def\bibfnamefont#1{#1}\fi
\expandafter\ifx\csname citenamefont\endcsname\relax
  \def\citenamefont#1{#1}\fi
\expandafter\ifx\csname url\endcsname\relax
  \def\url#1{\texttt{#1}}\fi
\expandafter\ifx\csname urlprefix\endcsname\relax\def\urlprefix{URL }\fi
\providecommand{\bibinfo}[2]{#2}
\providecommand{\eprint}[2][]{\url{#2}}

\bibitem[{\citenamefont{Olshanii}(1998)}]{CIR_Olshanii1998}
\bibinfo{author}{\bibfnamefont{M.}~\bibnamefont{Olshanii}},
  \bibinfo{journal}{Phys. Rev. Lett.} \textbf{\bibinfo{volume}{81}},
  \bibinfo{pages}{938} (\bibinfo{year}{1998}).

\bibitem[{\citenamefont{Bergeman et~al.}(2003)\citenamefont{Bergeman, Moore,
  and Olshanii}}]{CIR_Olshanii2003}
\bibinfo{author}{\bibfnamefont{T.}~\bibnamefont{Bergeman}},
  \bibinfo{author}{\bibfnamefont{M.~G.} \bibnamefont{Moore}}, \bibnamefont{and}
  \bibinfo{author}{\bibfnamefont{M.}~\bibnamefont{Olshanii}},
  \bibinfo{journal}{Phys. Rev. Lett.} \textbf{\bibinfo{volume}{91}},
  \bibinfo{pages}{163201} (\bibinfo{year}{2003}).

\bibitem[{\citenamefont{Paredes et~al.}(2004)\citenamefont{Paredes, Widera,
  Murg, Mandel, F{\"{o}}lling, Cirac, Shlyapnikov, H{\"{a}}nsch, and
  Bloch}}]{exp_TG_Paredes2004}
\bibinfo{author}{\bibfnamefont{B.}~\bibnamefont{Paredes}},
  \bibinfo{author}{\bibfnamefont{A.}~\bibnamefont{Widera}},
  \bibinfo{author}{\bibfnamefont{V.}~\bibnamefont{Murg}},
  \bibinfo{author}{\bibfnamefont{O.}~\bibnamefont{Mandel}},
  \bibinfo{author}{\bibfnamefont{S.}~\bibnamefont{F{\"{o}}lling}},
  \bibinfo{author}{\bibfnamefont{I.}~\bibnamefont{Cirac}},
  \bibinfo{author}{\bibfnamefont{G.~V.} \bibnamefont{Shlyapnikov}},
  \bibinfo{author}{\bibfnamefont{T.~W.} \bibnamefont{H{\"{a}}nsch}},
  \bibnamefont{and} \bibinfo{author}{\bibfnamefont{I.}~\bibnamefont{Bloch}},
  \bibinfo{journal}{Nature} \textbf{\bibinfo{volume}{429}},
  \bibinfo{pages}{277} (\bibinfo{year}{2004}).

\bibitem[{\citenamefont{Kinoshita}(2004)}]{exp_TG_Kinoshita2004}
\bibinfo{author}{\bibfnamefont{T.}~\bibnamefont{Kinoshita}},
  \bibinfo{journal}{Science} \textbf{\bibinfo{volume}{305}},
  \bibinfo{pages}{1125} (\bibinfo{year}{2004}).

\bibitem[{\citenamefont{Cazalilla et~al.}(2011)\citenamefont{Cazalilla, Citro,
  Giamarchi, Orignac, and Rigol}}]{RevModPhys.83.1405}
\bibinfo{author}{\bibfnamefont{M.~A.} \bibnamefont{Cazalilla}},
  \bibinfo{author}{\bibfnamefont{R.}~\bibnamefont{Citro}},
  \bibinfo{author}{\bibfnamefont{T.}~\bibnamefont{Giamarchi}},
  \bibinfo{author}{\bibfnamefont{E.}~\bibnamefont{Orignac}}, \bibnamefont{and}
  \bibinfo{author}{\bibfnamefont{M.}~\bibnamefont{Rigol}},
  \bibinfo{journal}{Rev. Mod. Phys.} \textbf{\bibinfo{volume}{83}},
  \bibinfo{pages}{1405} (\bibinfo{year}{2011}).

\bibitem[{\citenamefont{Guan et~al.}(2013)\citenamefont{Guan, Batchelor, and
  Lee}}]{GBL:2013}
\bibinfo{author}{\bibfnamefont{X.-W.} \bibnamefont{Guan}},
  \bibinfo{author}{\bibfnamefont{M.~T.} \bibnamefont{Batchelor}},
  \bibnamefont{and} \bibinfo{author}{\bibfnamefont{C.}~\bibnamefont{Lee}},
  \bibinfo{journal}{Rev. Mod. Phys.} \textbf{\bibinfo{volume}{85}},
  \bibinfo{pages}{1633} (\bibinfo{year}{2013}).

\bibitem[{\citenamefont{Girardeau}(1960)}]{BF_mapping_Girardeau1960}
\bibinfo{author}{\bibfnamefont{M.}~\bibnamefont{Girardeau}},
  \bibinfo{journal}{Journal of Mathematical Physics}
  \textbf{\bibinfo{volume}{1}}, \bibinfo{pages}{516} (\bibinfo{year}{1960}).

\bibitem[{\citenamefont{Giamarchi}(2004)}]{Giamarchi:2004}
\bibinfo{author}{\bibfnamefont{T.}~\bibnamefont{Giamarchi}},
  \bibinfo{journal}{(Oxford University Press, Oxford, 2004)}
  (\bibinfo{year}{2004}).

\bibitem[{\citenamefont{Korepin et~al.}(1993)\citenamefont{Korepin, Bogoliubov,
  and Izergin}}]{Korepin}
\bibinfo{author}{\bibfnamefont{V.~E.} \bibnamefont{Korepin}},
  \bibinfo{author}{\bibfnamefont{N.~M.} \bibnamefont{Bogoliubov}},
  \bibnamefont{and} \bibinfo{author}{\bibfnamefont{A.~G.}
  \bibnamefont{Izergin}}, \bibinfo{journal}{(Cambridge: Cambridge University
  Press)}  (\bibinfo{year}{1993}).

\bibitem[{\citenamefont{Takahashi}(1999)}]{Takahashi-b}
\bibinfo{author}{\bibfnamefont{M.}~\bibnamefont{Takahashi}},
  \bibinfo{journal}{(Cambridge: Cambridge University Press)}
  (\bibinfo{year}{1999}).

\bibitem[{\citenamefont{Essler et~al.}(2005)\citenamefont{Essler, Frahm, F, A,
  and E}}]{1D-Hubbard}
\bibinfo{author}{\bibfnamefont{F.~H.~L.} \bibnamefont{Essler}},
  \bibinfo{author}{\bibfnamefont{H.}~\bibnamefont{Frahm}},
  \bibinfo{author}{\bibfnamefont{G.}~\bibnamefont{F}},
  \bibinfo{author}{\bibfnamefont{K.}~\bibnamefont{A}}, \bibnamefont{and}
  \bibinfo{author}{\bibfnamefont{K.~V.} \bibnamefont{E}},
  \bibinfo{journal}{(Cambridge: Cambridge University Press)}
  (\bibinfo{year}{2005}).

\bibitem[{\citenamefont{Sutherland}(2004)}]{Sutherland-book}
\bibinfo{author}{\bibfnamefont{B.}~\bibnamefont{Sutherland}},
  \bibinfo{journal}{(Singapore: World Scientific)}  (\bibinfo{year}{2004}).

\bibitem[{\citenamefont{Lieb and Liniger}(1963)}]{BA_Lieb_Liniger1963}
\bibinfo{author}{\bibfnamefont{E.~H.} \bibnamefont{Lieb}} \bibnamefont{and}
  \bibinfo{author}{\bibfnamefont{W.}~\bibnamefont{Liniger}},
  \bibinfo{journal}{Phys. Rev.} \textbf{\bibinfo{volume}{130}},
  \bibinfo{pages}{1605} (\bibinfo{year}{1963}).

\bibitem[{\citenamefont{Yang}(1967)}]{BA_Yang1967}
\bibinfo{author}{\bibfnamefont{C.~N.} \bibnamefont{Yang}},
  \bibinfo{journal}{Phys. Rev. Lett.} \textbf{\bibinfo{volume}{19}},
  \bibinfo{pages}{1312} (\bibinfo{year}{1967}).

\bibitem[{\citenamefont{Jiang et~al.}(2015)\citenamefont{Jiang, Chen, and
  Guan}}]{Jiang:2015}
\bibinfo{author}{\bibfnamefont{Y.-Z.} \bibnamefont{Jiang}},
  \bibinfo{author}{\bibfnamefont{Y.-Y.} \bibnamefont{Chen}}, \bibnamefont{and}
  \bibinfo{author}{\bibfnamefont{X.-W.} \bibnamefont{Guan}},
  \bibinfo{journal}{Chin. Phys. B.} \textbf{\bibinfo{volume}{24}},
  \bibinfo{pages}{050311} (\bibinfo{year}{2015}).

\bibitem[{\citenamefont{Haller et~al.}(2009)\citenamefont{Haller, Gustavsson,
  Mark, Danzl, Hart, Pupillo, and Nagerl}}]{exp_sTG_Haller2009}
\bibinfo{author}{\bibfnamefont{E.}~\bibnamefont{Haller}},
  \bibinfo{author}{\bibfnamefont{M.}~\bibnamefont{Gustavsson}},
  \bibinfo{author}{\bibfnamefont{M.~J.} \bibnamefont{Mark}},
  \bibinfo{author}{\bibfnamefont{J.~G.} \bibnamefont{Danzl}},
  \bibinfo{author}{\bibfnamefont{R.}~\bibnamefont{Hart}},
  \bibinfo{author}{\bibfnamefont{G.}~\bibnamefont{Pupillo}}, \bibnamefont{and}
  \bibinfo{author}{\bibfnamefont{H.-C.} \bibnamefont{Nagerl}},
  \bibinfo{journal}{Science} \textbf{\bibinfo{volume}{325}},
  \bibinfo{pages}{1224} (\bibinfo{year}{2009}).

\bibitem[{\citenamefont{Yang et~al.}(2017)\citenamefont{Yang, Chen, Zheng, Sun,
  Dai, Guan, Yuan, and Pan}}]{Yang:2017}
\bibinfo{author}{\bibfnamefont{B.}~\bibnamefont{Yang}},
  \bibinfo{author}{\bibfnamefont{Y.-Y.} \bibnamefont{Chen}},
  \bibinfo{author}{\bibfnamefont{Y.-G.} \bibnamefont{Zheng}},
  \bibinfo{author}{\bibfnamefont{H.}~\bibnamefont{Sun}},
  \bibinfo{author}{\bibfnamefont{H.-N.} \bibnamefont{Dai}},
  \bibinfo{author}{\bibfnamefont{X.-W.} \bibnamefont{Guan}},
  \bibinfo{author}{\bibfnamefont{Z.-S.} \bibnamefont{Yuan}}, \bibnamefont{and}
  \bibinfo{author}{\bibfnamefont{J.-W.} \bibnamefont{Pan}},
  \bibinfo{journal}{Phys. Rev. Lett.} \textbf{\bibinfo{volume}{119}},
  \bibinfo{pages}{165701} (\bibinfo{year}{2017}).

\bibitem[{\citenamefont{Kinoshita et~al.}(2006)\citenamefont{Kinoshita, Wenger,
  and Weiss}}]{Kinoshita2006}
\bibinfo{author}{\bibfnamefont{T.}~\bibnamefont{Kinoshita}},
  \bibinfo{author}{\bibfnamefont{T.}~\bibnamefont{Wenger}}, \bibnamefont{and}
  \bibinfo{author}{\bibfnamefont{D.~S.} \bibnamefont{Weiss}},
  \bibinfo{journal}{Nature} \textbf{\bibinfo{volume}{440}},
  \bibinfo{pages}{900} (\bibinfo{year}{2006}).

\bibitem[{\citenamefont{Hofferberth et~al.}(2007)\citenamefont{Hofferberth,
  Lesanovsky, Fischer, Schumm, and Schmiedmayer}}]{Hofferberth2007}
\bibinfo{author}{\bibfnamefont{S.}~\bibnamefont{Hofferberth}},
  \bibinfo{author}{\bibfnamefont{I.}~\bibnamefont{Lesanovsky}},
  \bibinfo{author}{\bibfnamefont{B.}~\bibnamefont{Fischer}},
  \bibinfo{author}{\bibfnamefont{T.}~\bibnamefont{Schumm}}, \bibnamefont{and}
  \bibinfo{author}{\bibfnamefont{J.}~\bibnamefont{Schmiedmayer}},
  \bibinfo{journal}{Nature} \textbf{\bibinfo{volume}{449}},
  \bibinfo{pages}{324} (\bibinfo{year}{2007}).

\bibitem[{\citenamefont{Guan et~al.}(2007)\citenamefont{Guan, Batchelor, and
  Takahashi}}]{GBT:2007}
\bibinfo{author}{\bibfnamefont{X.-W.} \bibnamefont{Guan}},
  \bibinfo{author}{\bibfnamefont{M.~T.} \bibnamefont{Batchelor}},
  \bibnamefont{and}
  \bibinfo{author}{\bibfnamefont{M.}~\bibnamefont{Takahashi}},
  \bibinfo{journal}{Phys. Rev. A} \textbf{\bibinfo{volume}{76}},
  \bibinfo{pages}{043617} (\bibinfo{year}{2007}).

\bibitem[{\citenamefont{Deuretzbacher et~al.}(2014)\citenamefont{Deuretzbacher,
  Becker, Bjerlin, Reimann, and
  Santos}}]{effective_spin_chain_Deuretzbacher2014}
\bibinfo{author}{\bibfnamefont{F.}~\bibnamefont{Deuretzbacher}},
  \bibinfo{author}{\bibfnamefont{D.}~\bibnamefont{Becker}},
  \bibinfo{author}{\bibfnamefont{J.}~\bibnamefont{Bjerlin}},
  \bibinfo{author}{\bibfnamefont{S.~M.} \bibnamefont{Reimann}},
  \bibnamefont{and} \bibinfo{author}{\bibfnamefont{L.}~\bibnamefont{Santos}},
  \bibinfo{journal}{Phys. Rev. A} \textbf{\bibinfo{volume}{90}},
  \bibinfo{pages}{013611} (\bibinfo{year}{2014}).

\bibitem[{\citenamefont{Murmann et~al.}(2015)\citenamefont{Murmann,
  Deuretzbacher, Z\"urn, Bjerlin, Reimann, Santos, Lompe, and
  Jochim}}]{effective_spin_chain_Murmann2015}
\bibinfo{author}{\bibfnamefont{S.}~\bibnamefont{Murmann}},
  \bibinfo{author}{\bibfnamefont{F.}~\bibnamefont{Deuretzbacher}},
  \bibinfo{author}{\bibfnamefont{G.}~\bibnamefont{Z\"urn}},
  \bibinfo{author}{\bibfnamefont{J.}~\bibnamefont{Bjerlin}},
  \bibinfo{author}{\bibfnamefont{S.~M.} \bibnamefont{Reimann}},
  \bibinfo{author}{\bibfnamefont{L.}~\bibnamefont{Santos}},
  \bibinfo{author}{\bibfnamefont{T.}~\bibnamefont{Lompe}}, \bibnamefont{and}
  \bibinfo{author}{\bibfnamefont{S.}~\bibnamefont{Jochim}},
  \bibinfo{journal}{Phys. Rev. Lett.} \textbf{\bibinfo{volume}{115}},
  \bibinfo{pages}{215301} (\bibinfo{year}{2015}).

\bibitem[{\citenamefont{Levinsen et~al.}(2015)\citenamefont{Levinsen,
  Massignan, Bruun, and Parish}}]{effective_spin_chain_Levinsen2015}
\bibinfo{author}{\bibfnamefont{J.}~\bibnamefont{Levinsen}},
  \bibinfo{author}{\bibfnamefont{P.}~\bibnamefont{Massignan}},
  \bibinfo{author}{\bibfnamefont{G.~M.} \bibnamefont{Bruun}}, \bibnamefont{and}
  \bibinfo{author}{\bibfnamefont{M.~M.} \bibnamefont{Parish}},
  \bibinfo{journal}{Science Advances} \textbf{\bibinfo{volume}{1}},
  \bibinfo{pages}{e1500197} (\bibinfo{year}{2015}).

\bibitem[{\citenamefont{Yang et~al.}(2015)\citenamefont{Yang, Guan, and
  Pu}}]{effective_spin_chain_Puhan2015}
\bibinfo{author}{\bibfnamefont{L.}~\bibnamefont{Yang}},
  \bibinfo{author}{\bibfnamefont{L.}~\bibnamefont{Guan}}, \bibnamefont{and}
  \bibinfo{author}{\bibfnamefont{H.}~\bibnamefont{Pu}}, \bibinfo{journal}{Phys.
  Rev. A} \textbf{\bibinfo{volume}{91}}, \bibinfo{pages}{043634}
  (\bibinfo{year}{2015}).

\bibitem[{\citenamefont{Hu et~al.}(2016)\citenamefont{Hu, Pan, and
  Chen}}]{effective_spin_chain_Chenshu2016}
\bibinfo{author}{\bibfnamefont{H.}~\bibnamefont{Hu}},
  \bibinfo{author}{\bibfnamefont{L.}~\bibnamefont{Pan}}, \bibnamefont{and}
  \bibinfo{author}{\bibfnamefont{S.}~\bibnamefont{Chen}},
  \bibinfo{journal}{Phys. Rev. A} \textbf{\bibinfo{volume}{93}},
  \bibinfo{pages}{033636} (\bibinfo{year}{2016}).

\bibitem[{\citenamefont{Liu et~al.}(2017)\citenamefont{Liu, Chen, and
  Zhang}}]{effective_spin_chain_Liuyanxia2017}
\bibinfo{author}{\bibfnamefont{Y.}~\bibnamefont{Liu}},
  \bibinfo{author}{\bibfnamefont{S.}~\bibnamefont{Chen}}, \bibnamefont{and}
  \bibinfo{author}{\bibfnamefont{Y.}~\bibnamefont{Zhang}},
  \bibinfo{journal}{Phys. Rev. A} \textbf{\bibinfo{volume}{95}},
  \bibinfo{pages}{043628} (\bibinfo{year}{2017}).

\bibitem[{\citenamefont{Pan et~al.}(2017)\citenamefont{Pan, Liu, Hu, Zhang, and
  Chen}}]{effective_spin_chain_Panlei2017}
\bibinfo{author}{\bibfnamefont{L.}~\bibnamefont{Pan}},
  \bibinfo{author}{\bibfnamefont{Y.}~\bibnamefont{Liu}},
  \bibinfo{author}{\bibfnamefont{H.}~\bibnamefont{Hu}},
  \bibinfo{author}{\bibfnamefont{Y.}~\bibnamefont{Zhang}}, \bibnamefont{and}
  \bibinfo{author}{\bibfnamefont{S.}~\bibnamefont{Chen}},
  \bibinfo{journal}{Phys. Rev. B} \textbf{\bibinfo{volume}{96}},
  \bibinfo{pages}{075149} (\bibinfo{year}{2017}).

\bibitem[{\citenamefont{Tan}(2008{\natexlab{a}})}]{Tan2008_1}
\bibinfo{author}{\bibfnamefont{S.}~\bibnamefont{Tan}}, \bibinfo{journal}{Annals
  of Physics} \textbf{\bibinfo{volume}{323}}, \bibinfo{pages}{2952}
  (\bibinfo{year}{2008}{\natexlab{a}}).

\bibitem[{\citenamefont{Tan}(2008{\natexlab{b}})}]{Tan2008_2}
\bibinfo{author}{\bibfnamefont{S.}~\bibnamefont{Tan}}, \bibinfo{journal}{Annals
  of Physics} \textbf{\bibinfo{volume}{323}}, \bibinfo{pages}{2987}
  (\bibinfo{year}{2008}{\natexlab{b}}).

\bibitem[{\citenamefont{Tan}(2008{\natexlab{c}})}]{Tan2008_3}
\bibinfo{author}{\bibfnamefont{S.}~\bibnamefont{Tan}}, \bibinfo{journal}{Annals
  of Physics} \textbf{\bibinfo{volume}{323}}, \bibinfo{pages}{2971}
  (\bibinfo{year}{2008}{\natexlab{c}}).

\bibitem[{\citenamefont{Zhang and Leggett}(2009)}]{Zhang2009}
\bibinfo{author}{\bibfnamefont{S.}~\bibnamefont{Zhang}} \bibnamefont{and}
  \bibinfo{author}{\bibfnamefont{A.~J.} \bibnamefont{Leggett}},
  \bibinfo{journal}{Physical Review A} \textbf{\bibinfo{volume}{79}},
  \bibinfo{pages}{023601} (\bibinfo{year}{2009}).

\bibitem[{\citenamefont{Werner et~al.}(2009)\citenamefont{Werner, Tarruell, and
  Castin}}]{Werner2009}
\bibinfo{author}{\bibfnamefont{F.}~\bibnamefont{Werner}},
  \bibinfo{author}{\bibfnamefont{L.}~\bibnamefont{Tarruell}}, \bibnamefont{and}
  \bibinfo{author}{\bibfnamefont{Y.}~\bibnamefont{Castin}},
  \bibinfo{journal}{The European Physical Journal B}
  \textbf{\bibinfo{volume}{68}}, \bibinfo{pages}{401} (\bibinfo{year}{2009}).

\bibitem[{\citenamefont{Braaten and Platter}(2009)}]{contact_Braaten2009}
\bibinfo{author}{\bibfnamefont{E.}~\bibnamefont{Braaten}} \bibnamefont{and}
  \bibinfo{author}{\bibfnamefont{L.}~\bibnamefont{Platter}},
  \bibinfo{journal}{Laser Physics} \textbf{\bibinfo{volume}{19}},
  \bibinfo{pages}{550} (\bibinfo{year}{2009}).

\bibitem[{\citenamefont{Stewart et~al.}(2010)\citenamefont{Stewart, Gaebler,
  Drake, and Jin}}]{exp_contact_Jin2010}
\bibinfo{author}{\bibfnamefont{J.~T.} \bibnamefont{Stewart}},
  \bibinfo{author}{\bibfnamefont{J.~P.} \bibnamefont{Gaebler}},
  \bibinfo{author}{\bibfnamefont{T.~E.} \bibnamefont{Drake}}, \bibnamefont{and}
  \bibinfo{author}{\bibfnamefont{D.~S.} \bibnamefont{Jin}},
  \bibinfo{journal}{Phys. Rev. Lett.} \textbf{\bibinfo{volume}{104}},
  \bibinfo{pages}{235301} (\bibinfo{year}{2010}).

\bibitem[{\citenamefont{Yu et~al.}(2015)\citenamefont{Yu, Thywissen, and
  Zhang}}]{Contact_p_wave_Yu2015}
\bibinfo{author}{\bibfnamefont{Z.}~\bibnamefont{Yu}},
  \bibinfo{author}{\bibfnamefont{J.~H.} \bibnamefont{Thywissen}},
  \bibnamefont{and} \bibinfo{author}{\bibfnamefont{S.}~\bibnamefont{Zhang}},
  \bibinfo{journal}{Phys. Rev. Lett.} \textbf{\bibinfo{volume}{115}},
  \bibinfo{pages}{135304} (\bibinfo{year}{2015}).

\bibitem[{\citenamefont{Yoshida and Ueda}(2015)}]{contact_p_wave_Ueda2015}
\bibinfo{author}{\bibfnamefont{S.~M.} \bibnamefont{Yoshida}} \bibnamefont{and}
  \bibinfo{author}{\bibfnamefont{M.}~\bibnamefont{Ueda}},
  \bibinfo{journal}{Phys. Rev. Lett.} \textbf{\bibinfo{volume}{115}},
  \bibinfo{pages}{135303} (\bibinfo{year}{2015}).

\bibitem[{\citenamefont{Cui}(2016)}]{contact_p_wave_cui2016}
\bibinfo{author}{\bibfnamefont{X.}~\bibnamefont{Cui}}, \bibinfo{journal}{Phys.
  Rev. A} \textbf{\bibinfo{volume}{94}}, \bibinfo{pages}{043636}
  (\bibinfo{year}{2016}).

\bibitem[{\citenamefont{Cui and Dong}(2016)}]{contact_p_wave_cui2016_2}
\bibinfo{author}{\bibfnamefont{X.}~\bibnamefont{Cui}} \bibnamefont{and}
  \bibinfo{author}{\bibfnamefont{H.}~\bibnamefont{Dong}},
  \bibinfo{journal}{Phys. Rev. A} \textbf{\bibinfo{volume}{94}},
  \bibinfo{pages}{063650} (\bibinfo{year}{2016}).

\bibitem[{\citenamefont{Peng et~al.}(2016)\citenamefont{Peng, Liu, and
  Hu}}]{contact_p_wave_Peng2016}
\bibinfo{author}{\bibfnamefont{S.-G.} \bibnamefont{Peng}},
  \bibinfo{author}{\bibfnamefont{X.-J.} \bibnamefont{Liu}}, \bibnamefont{and}
  \bibinfo{author}{\bibfnamefont{H.}~\bibnamefont{Hu}}, \bibinfo{journal}{Phys.
  Rev. A} \textbf{\bibinfo{volume}{94}}, \bibinfo{pages}{063651}
  (\bibinfo{year}{2016}).

\bibitem[{\citenamefont{He et~al.}(2016)\citenamefont{He, Zhang, Chan, and
  Zhou}}]{contact_p_wave_Zhou2016}
\bibinfo{author}{\bibfnamefont{M.}~\bibnamefont{He}},
  \bibinfo{author}{\bibfnamefont{S.}~\bibnamefont{Zhang}},
  \bibinfo{author}{\bibfnamefont{H.~M.} \bibnamefont{Chan}}, \bibnamefont{and}
  \bibinfo{author}{\bibfnamefont{Q.}~\bibnamefont{Zhou}},
  \bibinfo{journal}{Phys. Rev. Lett.} \textbf{\bibinfo{volume}{116}},
  \bibinfo{pages}{045301} (\bibinfo{year}{2016}).

\bibitem[{\citenamefont{Luciuk et~al.}(2016)\citenamefont{Luciuk, Trotzky,
  Smale, Yu, Zhang, and Thywissen}}]{exp_contact_p_wave_Luciuk2016}
\bibinfo{author}{\bibfnamefont{C.}~\bibnamefont{Luciuk}},
  \bibinfo{author}{\bibfnamefont{S.}~\bibnamefont{Trotzky}},
  \bibinfo{author}{\bibfnamefont{S.}~\bibnamefont{Smale}},
  \bibinfo{author}{\bibfnamefont{Z.}~\bibnamefont{Yu}},
  \bibinfo{author}{\bibfnamefont{S.}~\bibnamefont{Zhang}}, \bibnamefont{and}
  \bibinfo{author}{\bibfnamefont{J.~H.} \bibnamefont{Thywissen}},
  \bibinfo{journal}{Nature Physics} \textbf{\bibinfo{volume}{12}},
  \bibinfo{pages}{1} (\bibinfo{year}{2016}).

\bibitem[{\citenamefont{Pricoupenko}(2008)}]{interaction_p_wave_Pricoupenko2008}
\bibinfo{author}{\bibfnamefont{L.}~\bibnamefont{Pricoupenko}},
  \bibinfo{journal}{Phys. Rev. Lett.} \textbf{\bibinfo{volume}{100}},
  \bibinfo{pages}{170404} (\bibinfo{year}{2008}).

\bibitem[{\citenamefont{Imambekov et~al.}(2010)\citenamefont{Imambekov,
  Lukyanov, Glazman, and Gritsev}}]{BA_p_wave_Imambekov2010}
\bibinfo{author}{\bibfnamefont{A.}~\bibnamefont{Imambekov}},
  \bibinfo{author}{\bibfnamefont{A.~A.} \bibnamefont{Lukyanov}},
  \bibinfo{author}{\bibfnamefont{L.~I.} \bibnamefont{Glazman}},
  \bibnamefont{and} \bibinfo{author}{\bibfnamefont{V.}~\bibnamefont{Gritsev}},
  \bibinfo{journal}{Phys. Rev. Lett.} \textbf{\bibinfo{volume}{104}},
  \bibinfo{pages}{040402} (\bibinfo{year}{2010}).

\bibitem[{\citenamefont{Hao et~al.}(2007)\citenamefont{Hao, Zhang, and
  Chen}}]{BA_p_wave_Hao2007}
\bibinfo{author}{\bibfnamefont{Y.}~\bibnamefont{Hao}},
  \bibinfo{author}{\bibfnamefont{Y.}~\bibnamefont{Zhang}}, \bibnamefont{and}
  \bibinfo{author}{\bibfnamefont{S.}~\bibnamefont{Chen}},
  \bibinfo{journal}{Phys. Rev. A} \textbf{\bibinfo{volume}{76}},
  \bibinfo{pages}{063601} (\bibinfo{year}{2007}).

\bibitem[{\citenamefont{Pan et~al.}(2018)\citenamefont{Pan, Chen, and
  Cui}}]{Pan:2018}
\bibinfo{author}{\bibfnamefont{L.}~\bibnamefont{Pan}},
  \bibinfo{author}{\bibfnamefont{S.}~\bibnamefont{Chen}}, \bibnamefont{and}
  \bibinfo{author}{\bibfnamefont{X.}~\bibnamefont{Cui}},
  \bibinfo{journal}{arXiv:1801.055902}  (\bibinfo{year}{2018}).

\bibitem[{\citenamefont{Chen et~al.}(2016)\citenamefont{Chen, Liu, and
  Hu}}]{Chen:2016}
\bibinfo{author}{\bibfnamefont{X.-L.} \bibnamefont{Chen}},
  \bibinfo{author}{\bibfnamefont{X.-J.} \bibnamefont{Liu}}, \bibnamefont{and}
  \bibinfo{author}{\bibfnamefont{H.}~\bibnamefont{Hu}}, \bibinfo{journal}{Phys.
  Rev. A} \textbf{\bibinfo{volume}{94}}, \bibinfo{pages}{033630}
  (\bibinfo{year}{2016}).

\bibitem[{\citenamefont{Olshanii and Dunjko}(2003)}]{Olshanii2003}
\bibinfo{author}{\bibfnamefont{M.}~\bibnamefont{Olshanii}} \bibnamefont{and}
  \bibinfo{author}{\bibfnamefont{V.}~\bibnamefont{Dunjko}},
  \bibinfo{journal}{Phys. Rev. Lett.} \textbf{\bibinfo{volume}{91}},
  \bibinfo{pages}{090401} (\bibinfo{year}{2003}).

\bibitem[{\citenamefont{Nandani et~al.}(2016)\citenamefont{Nandani,
  R{\"{o}}mer, Tan, and Guan}}]{correlation_Nandani2016}
\bibinfo{author}{\bibfnamefont{E.}~\bibnamefont{Nandani}},
  \bibinfo{author}{\bibfnamefont{R.~A.} \bibnamefont{R{\"{o}}mer}},
  \bibinfo{author}{\bibfnamefont{S.}~\bibnamefont{Tan}}, \bibnamefont{and}
  \bibinfo{author}{\bibfnamefont{X.~W.} \bibnamefont{Guan}},
  \bibinfo{journal}{New Journal of Physics} \textbf{\bibinfo{volume}{18}},
  \bibinfo{pages}{1} (\bibinfo{year}{2016}).

\bibitem[{\citenamefont{Girardeau et~al.}(2004)\citenamefont{Girardeau, Nguyen,
  and Olshanii}}]{Psuedopotential_p_wave_Girardeau2004}
\bibinfo{author}{\bibfnamefont{M.}~\bibnamefont{Girardeau}},
  \bibinfo{author}{\bibfnamefont{H.}~\bibnamefont{Nguyen}}, \bibnamefont{and}
  \bibinfo{author}{\bibfnamefont{M.}~\bibnamefont{Olshanii}},
  \bibinfo{journal}{Optics Communications} \textbf{\bibinfo{volume}{243}},
  \bibinfo{pages}{3} (\bibinfo{year}{2004}).

\bibitem[{\citenamefont{Grosse et~al.}(2004)\citenamefont{Grosse, Langmann, and
  Paufler}}]{Pseudopotential_p_wave_Grosse2004}
\bibinfo{author}{\bibfnamefont{H.}~\bibnamefont{Grosse}},
  \bibinfo{author}{\bibfnamefont{E.}~\bibnamefont{Langmann}}, \bibnamefont{and}
  \bibinfo{author}{\bibfnamefont{C.}~\bibnamefont{Paufler}},
  \bibinfo{journal}{Journal of Physics A: Mathematical and General}
  \textbf{\bibinfo{volume}{37}}, \bibinfo{pages}{4579} (\bibinfo{year}{2004}).

\bibitem[{\citenamefont{Yang and Yang}(1969)}]{Yang1969}
\bibinfo{author}{\bibfnamefont{C.~N.} \bibnamefont{Yang}} \bibnamefont{and}
  \bibinfo{author}{\bibfnamefont{C.~P.} \bibnamefont{Yang}},
  \bibinfo{journal}{Journal of Mathematical Physics}
  \textbf{\bibinfo{volume}{10}}, \bibinfo{pages}{1115} (\bibinfo{year}{1969}).

\bibitem[{\citenamefont{Bender et~al.}(2005)\citenamefont{Bender, Erker, and
  Granger}}]{Fermi_TG_Bender2005}
\bibinfo{author}{\bibfnamefont{S.~A.} \bibnamefont{Bender}},
  \bibinfo{author}{\bibfnamefont{K.~D.} \bibnamefont{Erker}}, \bibnamefont{and}
  \bibinfo{author}{\bibfnamefont{B.~E.} \bibnamefont{Granger}},
  \bibinfo{journal}{Phys. Rev. Lett.} \textbf{\bibinfo{volume}{95}},
  \bibinfo{pages}{230404} (\bibinfo{year}{2005}).

\bibitem[{\citenamefont{Girardeau and Minguzzi}(2006)}]{Fermi_TG_Girardeau2006}
\bibinfo{author}{\bibfnamefont{M.~D.} \bibnamefont{Girardeau}}
  \bibnamefont{and} \bibinfo{author}{\bibfnamefont{A.}~\bibnamefont{Minguzzi}},
  \bibinfo{journal}{Phys. Rev. Lett.} \textbf{\bibinfo{volume}{96}},
  \bibinfo{pages}{080404} (\bibinfo{year}{2006}).

\bibitem[{\citenamefont{Sekino et~al.}(2018)\citenamefont{Sekino, Tan, and
  Nishida}}]{contact_p_wave_Sekino2017}
\bibinfo{author}{\bibfnamefont{Y.}~\bibnamefont{Sekino}},
  \bibinfo{author}{\bibfnamefont{S.}~\bibnamefont{Tan}}, \bibnamefont{and}
  \bibinfo{author}{\bibfnamefont{Y.}~\bibnamefont{Nishida}},
  \bibinfo{journal}{Phys. Rev. A} \textbf{\bibinfo{volume}{97}},
  \bibinfo{pages}{013621} (\bibinfo{year}{2018}).

\end{thebibliography}

\end{document}